\documentclass[
preprintnumbers,a4paper, amsmath, amssymb, floatfix,
nofootinbib, twocolumn, wrapfloat byrevtex]{revtex4}

\usepackage{amsmath,mathrsfs,graphicx,epstopdf,placeins,float}
\usepackage{booktabs}
\usepackage{multirow}
\usepackage{pstricks}
\usepackage{tikz-feynman}
\usepackage{cancel}
\usetikzlibrary{arrows.meta}

\newcommand{\beq}[1]{\begin{equation}\label{#1}}
	\newcommand{\eeq}{\end{equation}}
\newcommand{\bea}[1]{\begin{eqnarray}\label{#1}}
	\newcommand{\eea}{\end{eqnarray}}

\begin{document} 
	
	\title{Direct Detection of Sub-GeV Dark Matter with $s>1$ through the DM-Electron Scattering (I): case study for $s=3/2$ and $s=2$ with (pseudo)scalar mediator}
	
	\author{Ke-yun Wu $^{a}$}
	\email{keyunwu@emails.bjut.edu.cn}
	\affiliation{\it ${}^a$ Faculty of Science, Beijing university of technology, Beijing, China}
	
	 \author{Acckin Lee $^{a}$}
	\email{lac@emails.bjut.edu.cn}
	\affiliation{\it ${}^a$ Faculty of Science, Beijing university of technology, Beijing, China}

	\begin{abstract}
	
	      The observable signal of higher-spin particles is rare in current particle collider experiments. In the meantime, there have no evident constraints on the spin of the dark matter particles. And thereby it is natural to consider the possibility that the higher-spin particles behaves as dark matter. In this work, we investigate the direct detection of (spin-$3/2$) gravitino dark matter and (spin-$2$) massive KK-graviton dark matter respectively through the DM-electron scattering. Both the spin-independent (SI) and spin-independent (SD) interactions of DM are considered. Using the experiment datas of XENON10, XENON100 and XENON1T, we could give the restriction relation between the DM mass and cross section. For simplification, we only compute the corss section at leading order, while merely focus on the (pseudo)scalar mediators. Moreover, by evaluating and analyzing the form factor of DM, we uncover that if the mediator is scalar there have no significant signals to distinguish the DM with spin $3/2,2$ from the one with spin $0,1/2,1$.
	    
		
	\end{abstract}
	\maketitle
	\section{Introduction}
	One of the major challenge in modern physics is how the the properties of dark matter can be detected\cite{Young:2016ala}. The dark matter is a hypothetical particle which do not emit or absorb radiation at any observable wavelengths and rarely interacts with baryons.~Due to 96\% matter of Universe's is dark matter and dark energy. Thus the detection of dark matter candidate and understanding of their elusive nature play an major role in astrophysics and particle physics. The attractive dark matter candidate is Weakly Interacting Massive Particles (WIMPs) with mass in a range $1\sim$ 100 GeV\cite{Dodelson:2003ft}. The WIMPs could interact with nucleus and produce observed nuclear recoils in direct detection experiment. However, the nuclear recoil energy cannot exceed the threshold of direct detection if dark matter mass is below 1 GeV. Such light dark matter could have enough  kinetic energy to induce electronic transitions\cite{Hochberg:2015pha,Essig:2015cda}. So the light dark matter can be detected by the electron recoil signal that is produced by dark matter-electron interaction
	\cite{Essig:2017kqs,Essig:2011nj,Essig:2012yx, XENON100:2011cza,Schutz:2016tid,Giudice:2017zke,Dolan:2017xbu,Dror:2020czw, Bloch:2020uzh,Flambaum:2020xxo,Guo:2021imc}.

	As electron mass is about half MeV, its recoils from colliding with the light dark matter may give a significant signatures in the detections. The dark matter-electron scattering \cite{Knapen:2021run,Hochberg:2021pkt} and the Migdal effect are potential technique to search for sub-GeV dark matter \cite{Flambaum:2020xxo,Baxter:2019pnz}. Currently, the dual-phase time projection chamber (TPC), liquid Xenon detector has given the most stringent direct-detection constraint on sub-GeV dark matter, especially the "S2" signal of the electrons which is drifted by electric fields upward into gaseous xenon when the dark matter particles scatter with the electron in the liquid. The resulting recoil electrons have sufficient energy to ionize other atoms. This signal includes the number of electrons that can be detected by photomultiplier tubes. In general, the experiment datas of XENON10, XENON100 and XENON1T are adopted to numerical analysis. 
	Most current dark matter interaction focus on an effective field theory with spin-independent (SI) scattering. As was proposed in \cite{Essig:2017kqs,Essig:2012yx,Essig:2011nj}, heavy mediator can make dark matter form factor to constant 1 and light mediator can make dark matter form factor to $1/q^{2}$ if the scattering is spin-independent (SI). For spin-dependent (SD) scattering, the form factor will be very complex due to the $\gamma_{\star}$ from mediator can cancel out the mass of target electron. As our previous work \cite{Wu:2022jln} explored, the limit of scattering cross section can be enhanced in SD scattering by effect of momentum transferred when dark matter spin is 0 and 1. We considered if the spin of dark matter can effect the limit of scattering cross section. So in this paper, we detect the sub-GeV dark matter with spin $s=3/2$ and $s=2$. 
	
	So far, in blocks of Standard Model(SM) and General Relativity(GR), we have known well with the spin-$0$ scalar particles, spin-$1/2$ spinor particles, spin-$1$ gauge bosons and spin-$2$ graviton. In supergravity, as the partner particle of graviton, the spin-$3/2$ particle (namely the gravitino) need to be introduced when local supersymmetry is considered. However, there has few observation signals about gravitino in experiments of high energy particle collider. If we turn our attention to the very early universe, as indicated by \cite{Kallosh:1999jj,Hashimoto:1998mu,Maroto:1999ch}, the amount of gravitinos can be produced considerably after inflation, even though the gravitational coupling is small. So, it is natural to view gravitino as one of potential candidates of dark matter which is highly secluded from the visible SM matters\cite{Steffen:2006hw,Rychkov:2007uq, Yu:2011by, Ding:2012sm, Ding:2013nvx, Chang:2017dvm, Garcia:2020hyo}. 
	
	On the other hand, due to the fact that DM can be only observed through the gravitational interactions, we have reason to conjecture that the DM particle is originated from a direct extension of the gravitational sector \cite{Babichev:2016bxi, Marzola:2017lbt, Aoki:2016zgp}. In other words, one of the most natural DM candidates is a massive graviton which could be involved to the gravitational sector in a cost-effective way. In particular, it is most straightforward way to add the massive graviton into the GR by constructing the bigravity theory \cite{Hassan:2011zd,Schmidt-May:2015vnx}. And then the multigravity theories are developed by \cite{Hinterbichler:2012cn,Hinterbichler:2011tt,deRham:2014zqa}. Alternatively, the graviton dark matter could also have an extra-dimensional origin \cite{Servant:2002aq,Cheng:2002ej,Feng:2003nr}. In models of universal extra dimensions \cite{Lykken:1996fj,Appelquist:2000nn}, both gravity and some SM fields could propagate in $D=4+d$ dimensions. As the extra dimension with a $Z_2$ symmetry is compactified, plenty of Kaluza-Klein (KK) particles are excited. Among these KK particles, only the lightest KK particle (LKP) is stable. Specifically, if the lightest KK graviton is assumed to be the LKP \cite{Feng:2003xh,Feng:2003uy}, and all KK particles originated by the SM fields ultimately decay to the KK graviton which becomes the only possible KK dark matter candidate.
	
	In summary, the candidates of DM can be the particles with spin-$0,\frac{1}{2},1,\frac{3}{2},2$, even the higher spin case $s>2$ \cite{Bellazzini:2019bzh}. So, one might wonder to know whether it is possible to distinguish the DM with different spins through the direct detection of the DM-Electron scattering. In \cite{Liu:2021avx}, by using the state-of-the-art many-body calculation method developed in area of atomic physics and the experimental data taken from XENON, the exclusion limit of SD DM-electron cross section at leading order are considered. Subsequently, as complementary studies, \cite{Wu:2022jln} extend the case of \cite{Liu:2021avx} into the scalar and vector DM with (psedo)scalar and (axil)vector mediators respectively, while the dependence of SD form factor on transferred momentum have been explored. Especially, in case of (pseudo)scalar mediator, there has no obvious distinction in direct detection of DM-Electron scattering for DMs with different spins. And thereby, in this paper, our main purpose is to generalize the situation of works \cite{Liu:2021avx}-\cite{Wu:2022jln} to the gravitino (spin-3/2) dark matter and graviton (spin-2) dark matter. Meanwhile, through the direct detection of the DM-Electron scattering, we aim to investigate that if there exists significant features for the dark matter with spin $s>1$ beyond the one with ordinary spin $s=0,\frac{1}{2},1$. For simplicity, in this work, we just consider the cross section $\bar{\chi}^s_{\text{DM}}\chi^s_{\text{DM}} \to \bar{e}e$ at tree-level, and merely focus on the (pseudo)scalar mediators.
	
	The paper is organized as follows. We introduce the dark matter model with $s=3/2$ and $s=2$ in Sec. \ref{sec2}, and then the result from numerical simulation is discussed in Sec. \ref{sec3}. The conclusion is given in Sec. \ref{sec4}. The trace operation of multiple $\gamma$-matrices product and the practical techniques of Clifford algebra are derived in detail in appendix A. The wave function and helicity sum for gravition are introduced in appendix B. At appendix C, we explicitly construct a phenomenological Lagrangian which describes the leading interactions between massive KK-graviton, radion and electron from the warped extra-dimensional model.
	

	\section{The theoretical calculation}\label{sec2}
	\subsection{The scattering rate of dark matter and electron}
	We follow the literature \cite{Essig:2017kqs, Essig:2012yx} to calculate the scattering rate between dark matter and electron. We treat the electron is single-particle states which can be described by RHF bound wave function \cite{Bunge:1993jsz}. The differential ionization rate that is caused by ionization electron is given by,
\begin{eqnarray}
 \frac{d R^{\rm ion}}{d  E_{R}}=
N\frac{\rho_{\chi}}{m_{\text{DM}}}\frac{d\langle\sigma^{nl}_{\rm ion} v\rangle}{d  E_R}
\label{eq4},
\end{eqnarray} 
in which, the local dark matter density is $\rho_{\chi}=0.4 \rm GeV/cm^{3}$  and $N=4.2\times 10^{27}$ \cite{Flambaum:2020xxo} is the number of the target Xenon. The $d\left<\sigma^{nl}_{\rm ion}v\right>$ is called as velocity-averaged differential ionization cross section. It can be written as 
\begin{eqnarray}
 \frac{d\langle\sigma^{nl}_{\rm ion} v\rangle}{d E_R}
&=&\frac{\bar{\sigma}_{e}}{8\mu_{\chi e}^{2}}\label{eq4}\\ \label{eqsg}
&\times&\int qdq \vert F_{\rm DM}(q)\vert^{2}
{\cal R}^{(\rm ion)}(E_R,q)\eta(v_{\rm min})\nonumber,
\end{eqnarray} 
twhere the $E_{R}$ and q are recoil energy and momentum of ionized electron, respectively. $\mu_{\text{DM} e}=m_{\text{DM}}m_{e}/(m_{\text{DM}}+m_{e})$ is called the reduced mass of dark matter ($m_{\text{DM}}$) and electron ($m_{e}$). The response function ${\cal R}^{\rm ion}$ is applied to encode the information of how the detector particles responds to the incoming dark matter particle\cite{Pandey:2018esq}.

For WIMP-nucleus scattering, the scattering cross section dependent on the form of interaction of dark matter with nucleons. In theories where WIMP primarily interact with nuclear spin, this interaction is called as spin-dependent. In this scenario, the scattering event will dependent on nuclear structure\cite{Agrawal:2010fh,Aalbers:2022dzr}. On the other hand, in theories where the scattering cross section is independent on the spin of nucleon, this interaction is called as sin-independent.
For light dark matter-electron scattering, the scattering cross section also sensitive to the form of interaction of light dark matter with electron. The electron wave function is not plan wave $e^{i q\cdot r}$ when the dark matter interact with electron spin, and it will be $e^{iq\cdot r}\sigma^{s}_{i,k}$\cite{Liu:2021avx}.

The response function of SI and SD scattering in Eq. \ref{eqsg} can be written as \cite{Liu:2021avx},
\begin{eqnarray}
 {\cal R}_{\rm SI}^{\rm ion}(E_R,q)&=&\sum_{J_{f}}\overline{\sum_{J_{i}}}
\left\vert\langle J_{f}\vert\sum_{i=1}^{Z}e^{i{\boldsymbol q}\cdot{\boldsymbol r}_{i}}\vert  
J_{i}\rangle\right\vert^{2}\nonumber\\
 && \times\delta(E_{J_{f}}-E_{J_{i}}-E_R).
 \label{eqRSI}
 \end{eqnarray}
 \begin{eqnarray}
 {\cal R}_{\rm SD}^{\rm ion}(E_R, q)&=&\sum_{J_{f}}\overline{\sum_{J_{i}}} \sum_{k}
\left\vert\langle J_{f}\vert\sum_{i=1}^{Z}e^{i{\boldsymbol q}\cdot{\boldsymbol r}_{i}}{\boldsymbol \sigma}_{i,k}^{s} 
\vert J_{i}\rangle\right\vert^{2}\nonumber\\
 &&\times\delta(E_{J_{f}}-E_{J_{i}}-E_R),
 \label{eqR}
 \end{eqnarray}
in which, the 4 $\times$ 4 Dirac spin matrices ${\boldsymbol \sigma}_{i,k}^{s}=\begin{pmatrix} \boldsymbol{\sigma} &0\\
0 & \boldsymbol{\sigma} \end{pmatrix}$ is applied to label the electron spin operator ${\boldsymbol{S}}_{e}$, where the $\boldsymbol{\sigma}$ is 2 $\times$ 2 Pauli matrices. The $E_{J_{f}}$, $E_{J_{i}}$ and $E_{R}$ is incoming energy, outgoing energy and recoiling energy, respectively. We assume the initial state and final state are $\vert J_{i}\rangle$ and $\langle J_{f}\vert$, respectively. We can get a ratio between spin-dependent and spin-independent response function by applying atom many-body method. The ratio is given by,
\begin{eqnarray}
 r=\frac{ {\cal R}_{\rm SD}^{\rm ion}}{{\cal R}_{\rm SI}^{\rm ion}}= 3.
\end{eqnarray}
In fact, the response function of SD (${\cal R}^{\rm ion}_{SD}$) is more complex than SI response function in relativity. It will dependent on recoil energy\cite{Liu:2021avx}. However, in our paper, the dark matter velocity is assumed non-relativity. So we can use this ratio to make the calculation of SD response function simpler.

In order to calculate the velocity distribution of dark matter, firstly, we assume the incoming velocity and outgoing momentum of dark matter is $v$ and $p_{\text{DM}}'$, so the momentum conservation requires
\begin{eqnarray}
\boldsymbol{q}=m_{\text{DM}}\boldsymbol{v}-\boldsymbol{p}_{\text{DM}}'=m_{e}\boldsymbol{v}_{e}
\end{eqnarray}
and  energy conservation requires
\begin{eqnarray}
    \Delta E_{e}=\frac {1}{2}m_{\text{DM}}v^{2}-\frac{\vert m_{\text{DM}}v-q\vert^{2}}{2m_{\text{DM}}}=\boldsymbol{v}\cdot\boldsymbol{q}-\frac {\boldsymbol{q}^2}{2m_{\text{DM}}}\label{energy}
\end{eqnarray}
in which, the $\Delta E_{e}$ is transferred energy of electron that dependent on binding energy $E_{b}$ and recoiling energy of electron $E_{R}$ . It requires $\Delta E_{e}=E_{b}+E_{R}$.
We can get the the minimum velocity $v_{\rm min}$ that required for ionizing electron from Eq. \ref{energy}
\begin{equation}
v_{\rm min}=\frac{|E_{\rm b}^{nl}|+E_{R}}{q}+\frac{q}{2m_{\text{DM}}}.
\label{eq3}
\end{equation} 
Then the velocity distribution of dark matter about the minimum velocity $\eta(v_{\rm min})$ that can be described by Maxwell-Boltzmann velocity distribution is given by
\begin{eqnarray}
  \eta(v_{\rm min})=\int_{v_{\rm min}} \frac{d^3 v}{v} f_\chi(v) \Theta (v-v_{\rm min}),
\end{eqnarray}
with
\begin{eqnarray}
 f_\chi(\vec{v}_\chi) \propto e^{-\frac{|\vec{v}_\chi +\vec{v}_E|^2}{v^2_0}} \Theta (v_{\rm esc}-|\vec{v}_{\chi}+\vec{v}_E|),
\end{eqnarray}
where we take the circular velocity $v_{0}=544$ km/s, the averaged Earth relative velocity $v_E=232$ km/s and the escape velocity $v_{esc}=544$ km/s~\cite{Smith:2006ym,Dehnen:1997cq}.

Next, we discuss the scattering cross section $\sigma_{e}$ in Eq. \ref{eqsg}. It can be written as,  
\begin{equation}
    \sigma_{e}=\bar{\sigma}_{e}\vert F_{\rm DM}(q)\vert^{2}
    \label{sigma1}
\end{equation}
where the momentum-dependent form factor $F_{\rm DM}(q)$ is defined by
\begin{equation}
    \vert F_{\rm DM}(q)\vert^{2}=\frac{\overline{\vert {\cal M}_{e}(q)\vert^{2}}}{\vert {\cal M}_{e}(q_{0})\vert^{2}}
    \label{form}
\end{equation}
and the reduced cross section $\bar{\sigma}_{e}$ is given by,
\begin{equation}
    \bar{\sigma}_{e}=\frac{\mu_{\chi e}^{2}\overline{\vert {\cal M}_{e}(q_{0})\vert^{2}}}{16\pi m_{\text{DM}}^{2}m_{e}^{2}}
\end{equation}
in which, $q_{0}=\alpha m_{e}$ is reference momentum. We can see that the form factor $F_{\rm DM}(q)$ dependent on the form of interaction of dark matter with electron. For contact interaction (heavy mediator), the form factor is a constant 1 ($F_{\rm DM}=1$). In this case, the transferred momentum will always less than mediator mass, so the effect of momentum can be ignored in propagator. However, for light mediator, the form factor will be more complicated with effect of transferred momentum. And the details will be discussed in next subsection.

\subsection{The gravitino as dark matter with scalar mediator}
A free massive spin-3/2 particle can be characterized by a vector-spinor $\psi_{\mu}$, whose dynamics are controlled by the Rarita-Schwinger Lagrangian \cite{Rarita:1941mf}
\begin{align}
\label{RaritaSchwinLagran}
&\mathcal{L}=-\frac{1}{2}\epsilon^{\mu\nu\rho\sigma}\bar{\Psi}_{\mu}\gamma_{\star}\gamma_{\nu}\partial_{\rho}\Psi_{\sigma}-\frac{1}{2}m_{3/2}\bar{\Psi}_{\mu}\gamma^{\mu\nu}\Psi_{\nu}
\end{align}
in which $\gamma_\star=\text{i}\gamma^0\gamma^1\gamma^2\gamma^3$ (It is usually denoted by $\gamma_5$ in textbook of QFT). Here, $\Psi_\mu$ is the Majorana spinor which satisfies the condition $\Psi_\mu=\Psi^C_\mu$ (see Appendix \ref{AppenSpinorConven}). From $\eqref{RaritaSchwinLagran}$, it is straightforward to derive the equation of motion for $\Psi_\mu$
\begin{align}
\label{EOMsForGravitino}
&\epsilon^{\mu\nu\rho\sigma}\gamma_{\star}\gamma_{\nu}\partial_{\rho}\Psi_{\sigma}+m_{3/2}\gamma^{\mu\nu}\Psi_{\nu}=0
\end{align}
In $\eqref{EOMsForGravitino}$, if we impose the constraint
\begin{align}
\label{GaugeFixGravitino}
&\gamma^\mu \Psi_\mu=0    
\end{align}
And then one could give rise to an extra constraint condition which is analogous to the Lorentz gauge for a spin-1 field (see Appendix.\ref{IntroToGravitino} for the derivation)
\begin{align}
\label{ContractRestrict}
&\partial^\mu \Psi_\mu=0
\end{align}
After inserting these constraints into the equation $\eqref{EOMsForGravitino}$, the equation of motion for $\Psi_\mu$ reduces to
\begin{align}
\label{DiracLikeGrano}
&\text{i}\gamma^\nu \partial_\nu\Psi_\mu -m_{3/2}\Psi_\mu=0
\end{align}
The constraints $\eqref{GaugeFixGravitino}$-$\eqref{ContractRestrict}$ allow us to reproduce the expected degree of freedom $(4-1-1)\times \frac{1}{2}\cdot2^{[4/2]}=4$ for massive spin-$3/2$ particle. In momentum space, the solution of $\eqref{GaugeFixGravitino}$-$\eqref{DiracLikeGrano}$ is constructed as
\begin{align}
&\hspace{-2.8mm}\Psi_{\mu}(x)\!=\!\sum_{\lambda}\!\int\!\!\frac{d^{3}\vec{p}}{2p^{0}(2\pi)^{3}}\big(\tilde{\Psi}_{\mu}^{(\lambda)}(\vec{p})e^{-ipx}\!\!+\!\tilde{\Psi}_{\mu}^{(\lambda)C}(\vec{p})e^{ipx}\big)\\
&\tilde{\psi}_{\mu}^{(\lambda)}(\vec{p})=\sum_{m,s}\delta_{\lambda,m+s}C_{\lambda;m,s}^{\frac{3}{2};1,\frac{1}{2}}u_{s}(\vec{p})\epsilon_{\mu}^{(m)}(\vec{p})
\end{align}
in which $C$ is the Clebsch-Gordan coefficient regarding to the addition of spin-$\frac{3}{2}$ particle with spin-$1$ particle. The more details about wave function $\tilde{\psi}_{\mu}^{(\lambda)}(\vec{p})$ and the corresponding helicity sum formula are given in Appendix.\ref{IntroToGravitino}.

We assume the $\Psi_\mu$ as dark matter, while choosing the real scalar particle as mediator. To simplify the computation, we consider the interaction
\begin{align}
\label{SIInteElecGravitino}
&\mathcal{L}_{\text{int}}\supset
g_{1}\phi\bar{\psi}_{\nu}\gamma^{\nu\mu}\psi_{\mu}+g_{2}\phi\bar{e}e
\end{align}
which corresponds to the spin-independent case. In term of the spin-dependent case, the interaction term $\eqref{SIInteElecGravitino}$ is modified to
\begin{align}
\label{SDInteElecGravitino}
&\mathcal{L}_{\text{int}}\supset
g_{1}\phi\bar{\psi}_{\nu}\gamma^{\nu\mu}\psi_{\mu}+g_{2}\phi\bar{e}\gamma_\star e  
\end{align}
According to $\eqref{SIInteElecGravitino}$-$\eqref{SDInteElecGravitino}$, the amplitudes for the scattering processes $\bar{\Psi}_\mu(p)  \Psi_\nu (q)\to \bar{\psi}_e(k) \psi_e(l)$ can be written 
\begin{align}
\nonumber
&\vert M_{\bar{\Psi}_{\mu}\Psi_{\nu}\to\bar{\psi}_{e}\psi_{e}}^{\text{SI}}\vert^{2}\!=\!\frac{1}{8}\frac{g_{1}^{2}g_{2}^{2}}{\big((q-p)^{2}-m_{\phi}^{2}\big)^{2}}\text{Tr}\big[(\cancel{k}+m_{\text{e}}I)(\cancel{l}+m_{\text{e}}I)\big]\\
\nonumber
&~~\times\!\text{Tr}\big[\gamma^{\alpha\beta}(\cancel{q}-m_{3/2}I)(A_{\beta\beta^{\prime}}I-\frac{1}{3}A_{\beta\rho}A_{\beta^{\prime}\lambda}\gamma^{\rho}\gamma^{\lambda})\\
\label{SquaAmplitudeSpin3d2SI}
&~~\cdot\gamma^{\beta^{\prime}\alpha^{\prime}}(\cancel{p}-m_{3/2}I)(A_{\alpha^{\prime}\alpha}I-\frac{1}{3}A_{\alpha^{\prime}\rho^{\prime}}A_{\alpha\lambda^{\prime}}\gamma^{\rho^{\prime}}\gamma^{\lambda^{\prime}})\big]\\
\nonumber
&\hspace{-5mm}\vert M_{\bar{\Psi}_{\mu}\Psi_{\nu}\to\bar{\psi}_{e}\psi_{e}}^{\text{SD}}\vert^{2}\!=\!\frac{1}{8}\frac{g_{1}^{2}g_{2}^{2}}{\big((q-p)^{2}-m_{\phi}^{2}\big)^{2}}\text{Tr}\big[\gamma_\star(\cancel{k}+m_{\text{e}}I)\gamma_\star(\cancel{l}+m_{\text{e}}I)\big]\\
\nonumber
&~~\times\!\text{Tr}\big[\gamma^{\alpha\beta}(\cancel{q}-m_{3/2}I)(A_{\beta\beta^{\prime}}I-\frac{1}{3}A_{\beta\rho}A_{\beta^{\prime}\lambda}\gamma^{\rho}\gamma^{\lambda})\\
\label{SquaAmplitudeSpin3d2SD}
&~~\cdot\gamma^{\beta^{\prime}\alpha^{\prime}}(\cancel{p}-m_{3/2}I)(A_{\alpha^{\prime}\alpha}I-\frac{1}{3}A_{\alpha^{\prime}\rho^{\prime}}A_{\alpha\lambda^{\prime}}\gamma^{\rho^{\prime}}\gamma^{\lambda^{\prime}})\big]
\end{align}
where the symbol $A_{\mu\nu}$ denotes $A_{\mu\nu}=\eta_{\mu\nu}-\frac{q_\mu q_\nu}{m^2_{\text{3/2}}}$. Basing on the technique in dealing with the trace of multiple $\gamma$-matrices product provided in Appendix.\ref{AppenSpinorConven}, we obtain
\begin{align}
&\vert M_{\bar{\Psi}_{\mu}\Psi_{\nu}\to\bar{\psi}_{e}\psi_{e}}^{\text{SI}}\vert^{2}=\frac{g_{\chi}^2 g_{e}^2}{2(t-m_{a}^{2})^2}\big(-\frac{t}{2}+2m_{e}^{2}\big)\\
\nonumber
&\times\!\big(28m_{3/2}^{2}\!+\!\frac{(2m_{3/2}^{2}-t)^{2}}{m_{3/2}^{2}}\!-\!\frac{1}{9}\big(196m_{3/2}^{2}\!-\!\frac{53(2m_{3/2}^{2}-t)^{2}}{m_{3/2}^{2}}\big)\big)
\end{align}
and 
\begin{align}
&\vert M_{\bar{\Psi}_{\mu}\Psi_{\nu}\to\bar{\psi}_{e}\psi_{e}}^{\text{SD}}\vert^{2}=\frac{g_{\chi}^2 g_{e}^2}{2(t-m_{a}^{2})^2}\left(\frac{t}{2}\right)\\
\nonumber
&\times\!\big(28m_{3/2}^{2}\!+\!\frac{(2m_{3/2}^{2}-t)^{2}}{m_{3/2}^{2}}\!-\!\frac{1}{9}\big(196m_{3/2}^{2}\!-\!\frac{53(2m_{3/2}^{2}-t)^{2}}{m_{3/2}^{2}}\big)\big)    
\end{align}
The momentum-dependent form factor $F_{\rm DM}$ of spin-independent and spin-dependent scattering in Eq.  \eqref{form} can be written as 

	\begin{eqnarray}
	\vert F_{\rm DM}^{\rm SI}\vert^{2}=\frac{(4m_{e}^{2}+q^{2})(m_{a}^{2}+q_{0}^{2})^{2}}{(4m_{e}^{2}+q_{0}^{2})(m_{a}^{2}+q^{2})^{2}}\times I_{3/2}
	\label{ifdm}
	\end{eqnarray}
	and 
\begin{eqnarray}
	\vert F_{\rm DM}^{\rm SD}\vert^{2}=\frac{q^{2}(m_{a}^{2}+q_{0}^{2})^{2}}{q_{0}^{2}(m_{a}^{2}+q^{2})^{2}}\times D_{3/2}
	\label{dfdm}
	\end{eqnarray}
	where,
	\begin{eqnarray}
	I_{3/2}=D_{3/2}=\frac{(124m_{3/2}^{2}q^{2}+152m_{3/2}^{4}+31q^{4})}{(124m_{3/2}^{2}q_{0}^{2}+152m_{3/2}^{4}+31q_{0}^{4})}.
	\end{eqnarray}
	
\subsection{The massive graviton as dark matter with scalar mediator}

The dynamics of a free massive spin-2 particle is described by the Fierz-Pauli action
\begin{align}
\nonumber
&S\!=\!\int d^{4}x~\big\{\frac{1}{2}h\Box h-\frac{1}{2}h^{\mu\nu}\Box h_{\mu\nu}-h\partial_{\mu}\partial_{\nu}h^{\mu\nu}\\
\label{spin2FPaction}
&\quad\quad\quad\quad-\partial_{\mu}h^{\mu\nu}\partial_{\rho}h_{~\nu}^{\rho}-\frac{1}{2}m^{2}(h^{\mu\nu}h_{\mu\nu}-h^{2})
\end{align}
in which $\Box=\partial^\mu \partial_\mu$ is the d'Alembert operator. The corresponding equation of motion is
\begin{align}
\nonumber
&0=(\Box+m^{2})(\eta_{\mu\nu}h-h_{\mu\nu})+2\partial_{(\mu}\partial^{\alpha}h_{\alpha\nu)}\\
\label{EOMsGraviton}
&~~-\partial_{\alpha}\partial_{\beta}(\eta_{\mu\nu}h^{\alpha\beta}+\delta_{\mu}^{\alpha}\delta_{\nu}^{\beta}h)
\end{align}
In order to reproduce the correct degrees of freedom, one could impose the traceless condition
\begin{align}
\label{TracelessCondition}
&h=\eta^{\mu\nu}h_{\mu\nu}=0
\end{align}
Taking $\eqref{TracelessCondition}$ back to $\eqref{EOMsGraviton}$ and contracting the resulting equation, the transverse polarization condition is implied
\begin{align}
\label{TransverseCon}
&\partial^\mu h_{\mu\nu}=0
\end{align}
After plugging $\eqref{TracelessCondition}$ and $\eqref{TransverseCon}$ into the equations $\eqref{EOMsGraviton}$, we acquire the standard wave equation
\begin{align}
&(\Box+m^2)h_{\mu\nu}=0
\end{align}
Meanwhile, the constraints $\eqref{TracelessCondition}$-$\eqref{TransverseCon}$ help us generate the expected degree of freedom $10-1-4=5$ for massive spin-$2$ particle. 

Here, the massive graviton $h_{\mu\nu}$ is supposed to be the DM candidate. Besides, the scattering process $hh\to\bar{e}e$ at leading order is implemented by the interactions
\begin{align}
\label{InteBetElecGraSI}
&\hspace{-2.3mm}\mathcal{L}_{\text{int}}\!=\!g_{1}\boldsymbol{r}(h^{\mu\nu}h_{\mu\nu}\!\!-\!\!h^{2})\!\!-\!\!g_{2}\boldsymbol{r}\big(\text{i}\bar{\psi}_{\text{e}}\gamma^{\mu}\partial_{\mu}\psi_{\text{e}}\!\!-\!\!\frac{4}{3}m_{\text{e}}\bar{\psi}_{\text{e}}\psi_{\text{e}}\big)
\end{align}
where the massive scalar field $\boldsymbol{r}$ represents the radion which is produced from the scalar perturbations on the extra dimension. The details about the derivation of interaction term $\eqref{InteBetElecGraSI}$ from the warped extra-dimensional model  can be found in Appendix \ref{EffActionFromBrane}. In spin-dependent case, the interaction $\eqref{InteBetElecGraSI}$ is generalized to
\begin{align}
\label{InteBetElecGraSD}
&\hspace{-5.96mm}\mathcal{L}_{\text{int}}\!\!=\!\!g_{1}\boldsymbol{r}(h^{\mu\nu}h_{\mu\nu}\!\!-\!\!h^{2})\!\!-\!\!g_{2}\boldsymbol{r}\big(\text{i}\bar{\psi}_{\text{e}}\gamma^{\mu}\gamma_\star\partial_{\mu}\psi_{\text{e}}\!\!-\!\!\frac{4}{3}m_{\text{e}}\bar{\psi}_{\text{e}}\gamma_\star\psi_{\text{e}}\big)
\end{align}
From $\eqref{InteBetElecGraSI}$-$\eqref{InteBetElecGraSD}$, the amplitudes for the $h(p)h(q)\to \bar{e}(k) e(l)$ at leading order take the expressions
\begin{align}
\nonumber
&\hspace{-2mm}\vert M_{hh\to\bar{e}e}^{\text{SI}}\vert^{2}\!=\!\frac{g_{1}^{2}g_{2}^{2}}{10}\frac{\text{Tr}\big((\cancel{k}\!+\!m_{e}I)(\frac{4}{3}\text{i}m_{e}I\!+\!\text{i}\cancel{l})(\cancel{l}\!+\!m_{e}I)(\frac{4}{3}\text{i}m_{e}I\!-\!\text{i}\cancel{l})\big)}{\big((l-k)^{2}-m_{\boldsymbol{r}}^{2}\big)^{2}}\\
\nonumber
&\quad\quad\quad~\cdot\frac{1}{2}\mathcal{B}_{\mu\nu\mu^{\prime}\nu^{\prime}}(q)(\eta^{\alpha^{\prime}\mu^{\prime}}\eta^{\beta^{\prime}\nu^{\prime}}\!+\!\eta^{\alpha^{\prime}\nu^{\prime}}\eta^{\beta^{\prime}\mu^{\prime}}\!-\!2\eta^{\alpha^{\prime}\beta^{\prime}}\eta^{\mu^{\prime}\nu^{\prime}})\\
\label{hhTOeeSI}
&\quad\quad\quad~\cdot\frac{1}{2}\mathcal{B}_{\alpha\beta\alpha^{\prime}\beta^{\prime}}(p)(\eta^{\alpha\mu}\eta^{\beta\nu}\!+\!\eta^{\alpha\nu}\eta^{\beta\mu}\!-\!2\eta^{\alpha\beta}\eta^{\mu\nu})
\end{align}
\begin{align}
\nonumber
&\hspace{-6mm}\vert M_{hh\to\bar{e}e}^{\text{SD}}\vert^{2}\!\!=\!\!-\frac{g_{1}^{2}g_{2}^{2}}{10}\frac{\text{Tr}\big((\cancel{k}\!+\!m_{e}I)(\frac{4}{3}\text{i}m_{e}\!+\!\text{i}\cancel{l})\gamma_{\star}(\cancel{l}\!+\!m_{e}I)(\frac{4}{3}\text{i}m_{e}\!+\!\text{i}\cancel{l})\gamma_{\star}\big)}{\big((l-k)^{2}-m_{\text{r}}^{2}\big)^{2}}  \\
\nonumber
&\quad\quad\quad\cdot\frac{1}{2}\mathcal{B}_{\mu\nu\mu^{\prime}\nu^{\prime}}(q)(\eta^{\alpha^{\prime}\mu^{\prime}}\eta^{\beta^{\prime}\nu^{\prime}}+\eta^{\alpha^{\prime}\nu^{\prime}}\eta^{\beta^{\prime}\mu^{\prime}}-2\eta^{\alpha^{\prime}\beta^{\prime}}\eta^{\mu^{\prime}\nu^{\prime}})\\
&\quad\quad\quad\cdot\frac{1}{2}\mathcal{B}_{\alpha\beta\alpha^{\prime}\beta^{\prime}}(p)(\eta^{\alpha\mu}\eta^{\beta\nu}\!+\!\eta^{\alpha\nu}\eta^{\beta\mu}\!-\!2\eta^{\alpha\beta}\eta^{\mu\nu})
\end{align}
in which the notation $\mathcal{B}_{\alpha\beta\mu\nu}(p)$ represents
\begin{align}
\nonumber
&\mathcal{B}_{\alpha\beta\mu\nu}(p)\!=\!\mathcal{G}_{\alpha\mu}(p)\mathcal{G}_{\beta\nu}(p)\!+\!\mathcal{G}_{\alpha\nu}(p)\mathcal{G}_{\mu\beta}(p)\!-\!\frac{2}{3}\mathcal{G}_{\alpha\beta}(p)\mathcal{G}_{\mu\nu}(p)\\
\label{hhTOeeSD}
&\mathcal{G}_{\alpha\beta}(p)=\eta_{\alpha\beta}-\frac{p_{\alpha}p_{\beta}}{m_{h}^{2}}
\end{align}
In the light of the techniques provided in Appendix.\ref{AppenSpinorConven} and Appendix.\ref{EffActionFromBrane}, the amplitude $\eqref{hhTOeeSI}$ and $\eqref{hhTOeeSD}$ reduce to
\begin{align}
&\vert M_{hh\to\bar{e}e}^{\text{SI}}\vert^{2}\!=\!\frac{g_1^{2}g_2^{2}}{10(t-m_{\text{r}}^2)^{2}}\big(-\frac{7}{18}m_{e}^{2}t+\frac{14}{9}m_{e}^{4}\big)\\
\nonumber
&\times\left(\frac{-2300}{9}+\frac{22}{9}\left(\frac{(2m_{h}^2-t)^2}{m_{h}^{4}}\right)+\frac{29}{9}\left(\frac{(2m_{h}^{2}-t)^{4}}{128m_{h}^{8}}\right)\right)
\end{align}
and
\begin{align}
&\vert M_{hh\to\bar{e}e}^{\text{SD}}\vert^{2}\!\!=\!\!\frac{g_1^{2}g_2^{2}}{10(t-m_{\text{r}}^2)^{2}}\left(\frac{7}{18}m_{e}^{2}t\right)\\
\nonumber
&\times\left(\frac{-2300}{9}+\frac{22}{9}\left(\frac{(2m_{h}^2-t)^2}{m_{h}^{4}}\right)+\frac{29}{9}\left(\frac{(2m_{h}^{2}-t)^{4}}{128m_{h}^{8}}\right)\right)
\end{align}

The momentum-dependent form factor $F_{\rm DM}$ of spin-independent and spin-dependent scattering in Eq.  \eqref{form} can be written as 

	\begin{eqnarray}
	\vert F_{\rm DM}^{\rm SI}\vert^{2}=\frac{(4m_{e}^{2}+q^{2})(m_{r}^{2}+q_{0}^{2})^{2}}{(4m_{e}^{2}+q_{0}^{2})(m_{r}^{2}+q^{2})^{2}}\times I_{2}
	\label{ifdm}
	\end{eqnarray}
	and 
\begin{eqnarray}
	\vert F_{\rm DM}^{\rm SD}\vert^{2}=\frac{q^{2}(m_{r}^{2}+q_{0}^{2})^{2}}{q_{0}^{2}(m_{r}^{2}+q^{2})^{2}}\times D_{2}
	\label{dfdm}
	\end{eqnarray}
	where,
	\begin{eqnarray}
	I_{2}=D_{2}\approx\frac{12192m_{h}^{6}q^{2}+3512m_{h}^{4}q^{4}+232m_{h}^{2}q^{6}}{12192m_{h}^{6}q_{0}^{2}+3512m_{h}^{4}q_{0}^{4}+232m_{h}^{2}q_{0}^{6}}.
	\end{eqnarray}

\section{Numerical analysis and discussion}	\label{sec3}
We use the procedure from \cite{Essig:2012yx} to simulate the constrains of scattering cross section. In this procedure, the minimum energy that can produce a quantum number is 13.8ev which is labeled as $W$. So we can know that the quantum number which is produced by electron recoil energy $E_{\rm R}$ is $n^{(1)} = {\rm Floor}~(E_{\rm R}/W)$. The quantum number $n^{1}$ include electron number $n_{e}$ and scintillation photons number $n_{\gamma}$.~In which, we set $f_{R}=0$, $f_{e}=0.83$ as experimental fiducial valves.
The electron in following shells: $(5p,~5s,~4d,~4p,~4s)$ are considered, whose binding energies correspond to $(12.4,~25.7,~75.6,~163.5,~213.8)$ ev. Due to photoionization effect, the photons can also produce some additional quantum numbers $n^{(2)}= (5p,~5s,~4d,~4p,~4s)= (0,~0,~4,~6-10,~3-15)$  \cite{Essig:2017kqs,Cao:2020bwd}. We listed binding energy and additional quanta numbers from different shells in Tab. \ref{tabqe}. We can see that the total quantum number is $(n^{(1)}+n^{(2)})$ which is constituted by observed electron $n_{e}$ and scintillation photons $n_{\gamma}$. The number of electron $n_{e}$ and the number of photon $n_{\gamma}$ apply a binomial distribution.
\begin{table}[]
	\begin{tabular}{| l || l | l | l | l | l |}
		\hline 
		shell& 5$p^{6}$ & 5$s^{2}$ & 4$d^{10}$ & 4$p^{6}$ & 4 $s^{2}$\\
		\cline{1-6}
		Binding Energy[eV] & 12.6 & 25.7 & 75.6 & 163.5 & 213.8\\	
		\cline{1-6}
		Photon Energy[ev] &\quad-& 13.3 & 63.2 & 87.9 & 201.4\\
		\cline{1-6}
		Additional Quanta & 0 & 0 & 4 & 6-10 & 3-15\\
		\cline{1-6}
		\hline
	\end{tabular}
\caption{The number of additional quanta and binding energy 
from the Xenon ($5p^{6},~5s^{2},~4d^{10},~4p^{6},~4s^{2}$) shells. }\label{tabqe}
\end{table}

On experimental side, the weak signal of electron will be amplified to photoelectrons (PE) by photomultipier tuber. An event of $n_{e}$ can produce PE numbers which obey gaussian distribution. In this distribution, we define parameter mean $n_{e}\mu$ and width $\sqrt{n_{e}\sigma}$, where $\mu = 27~(19.7, 11.4)$ and $\sigma = 6.7~ (6.2, 2.8)$ for XENON10 (XENON100, XENON1T). We follow the Ref. \cite{Essig:2012yx,XENON100:2011cza,XENON:2019gfn} to define the PE bin. For XENON 10 (15kg-days), we set seven PE bin which is $b_{1}=$[14,41], $b_{2}=$[41,68], $b_{3}=$[68,95], $b_{4}=$[95,122], $b_{5}=$[122,147], $b_{6}=$[149,176],$b_{7}=$[176,203], respectively. For XENON100 (30kg-days), we set six PE bin which is $b_{1}=$[80,90], $b_{2}=$[90,110], $b_{3}=$[110,130], $b_{3}$=[130,150], $b_{4}=$[150,170], $b_{5}=$[170,190], $b_{6}=$[190,210], respectively. The XENON1T (1.5tones-yaers) is very specially, we just set one PE bin which is [165,275].

\begin{figure}[H]
\centering
\includegraphics[scale=0.58]{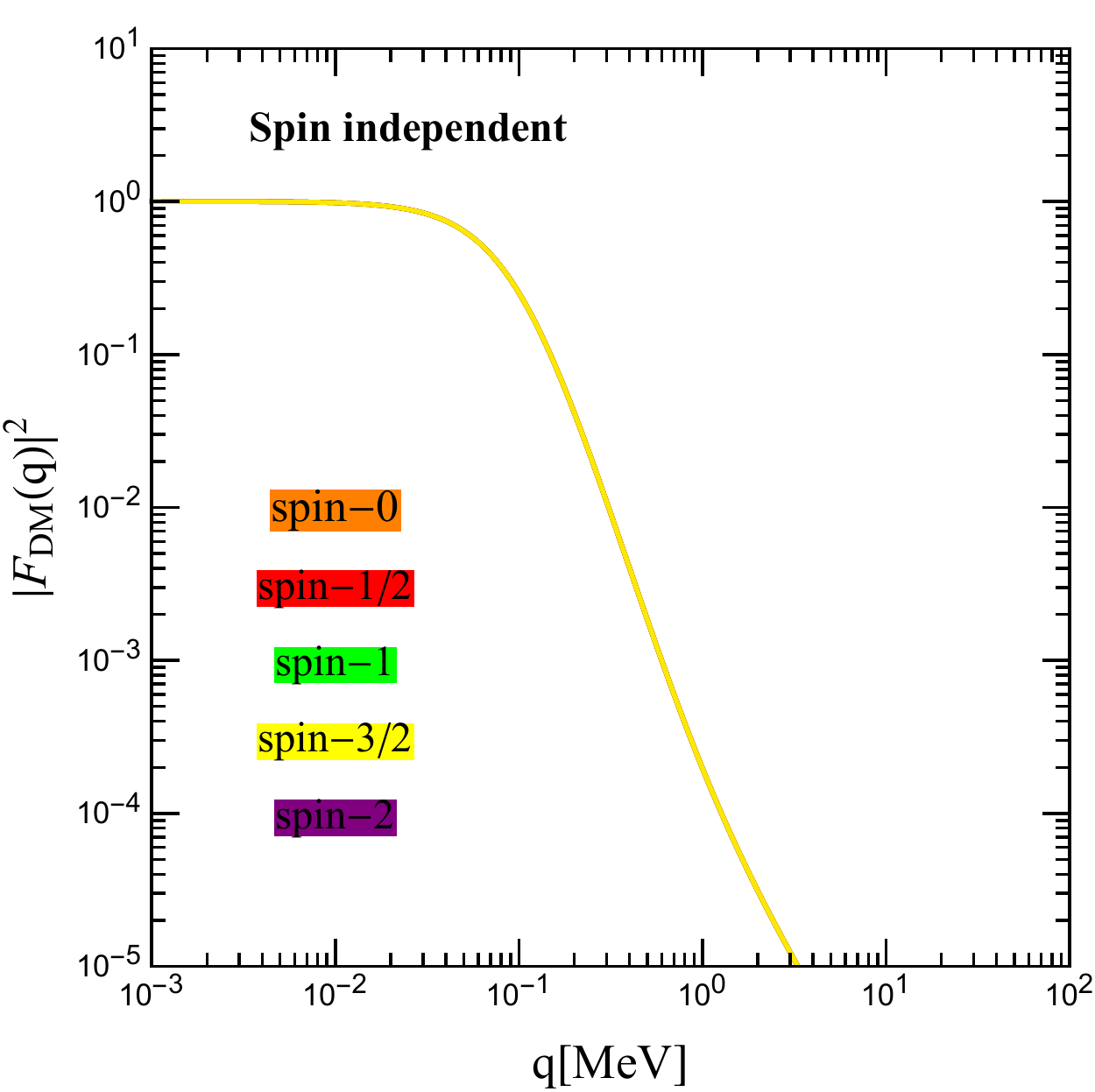}
\includegraphics[scale=0.58]{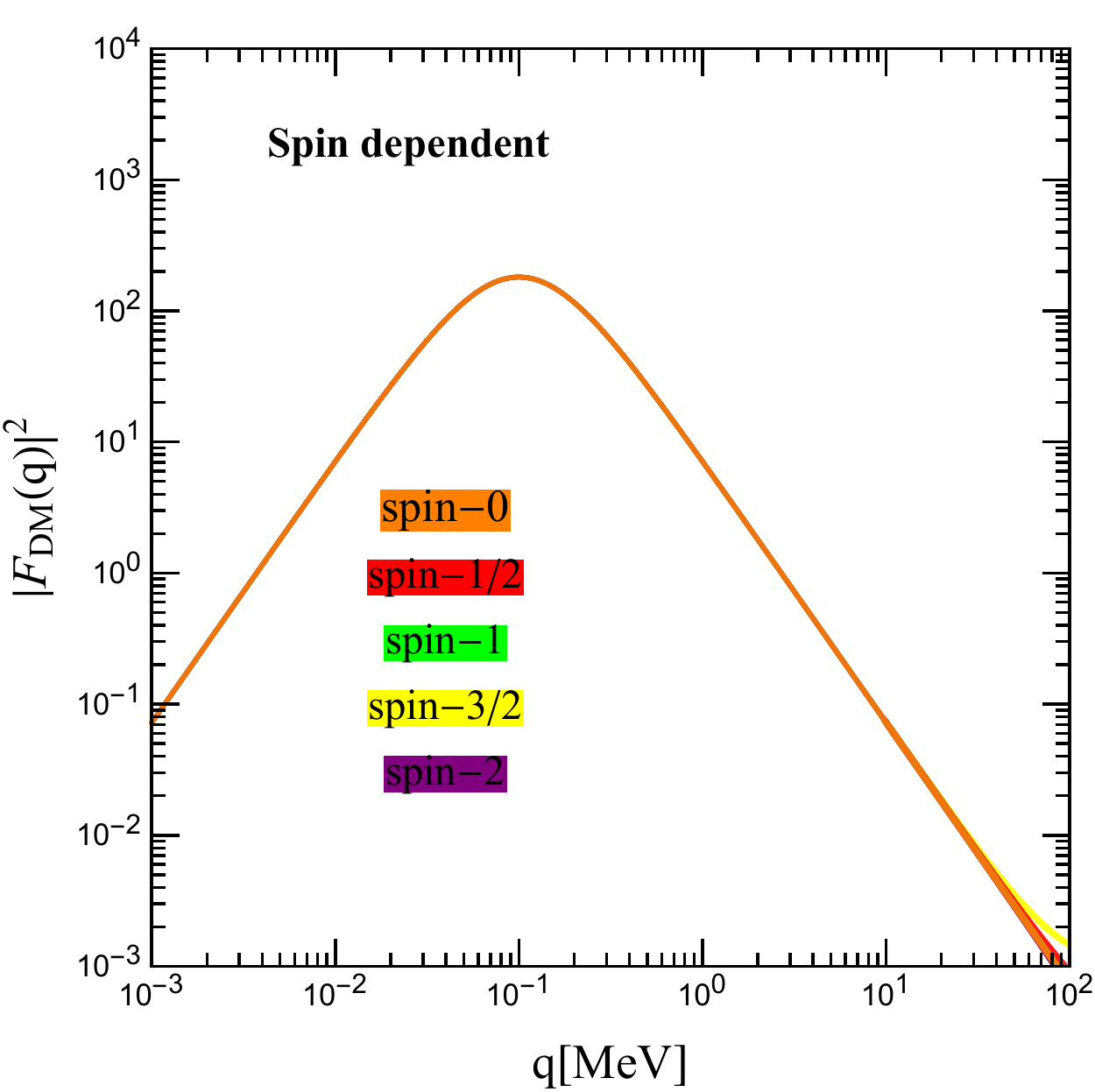}
\caption{The relation of form factor $\vert F_{\rm DM}(q)\vert^{2}$ and momentum transfer $q$ from different dark matter spin are shown. The left picture show the Spin-independent scattering and the right picture show the Spin-dependent scattering. In which, we set the mass of mediator is 0.5MeV and the dark matter mass is 100MeV.}	
\label{fdm}
\end{figure}

In our previous paper\cite{Wu:2022jln}, the form factor of scalar and vector dark matter scattering with electron at scalar mediator have been calculated. The form factor with $s=1/2$ is calculated in literature \cite{Liu:2021avx}. We find that the form factor of $s=0,~1/2,~1$ have the same expression with $s=3/2,~2$ except $I(D)$, in which, the $I_{0}=D_{0}=1$, $I_{1/2}=D_{1/2}=(4m_{\text{DM}}^{2}+q^{2})/(4m_{\text{DM}}^{2}+q_{0}^{2})$ and $I_{1}=D_{1}=(4m_{\text{DM}}^{2}q^{2}+12m_{\text{DM}}^{4}+q^{4})/(4m_{\text{DM}}^{2}q_{0}^{2}+12m_{\text{DM}}^{4}+q_{0}^{4})$. We show the dependence of dark matter form factor on transferring momentum q at $m_{a}(m_{r})=0.1$ MeV and $m_{\text{DM}}=100$ MeV in Fig. \ref{fdm}. It can be seen that the general behavior of dark matter form factor is that it will have same result with dark matter spin $s=0,~1/2,~1,~3/2,~2$ whether SI scattering or SD scattering. 

\begin{figure}[H]
\centering
\includegraphics[scale=0.60]{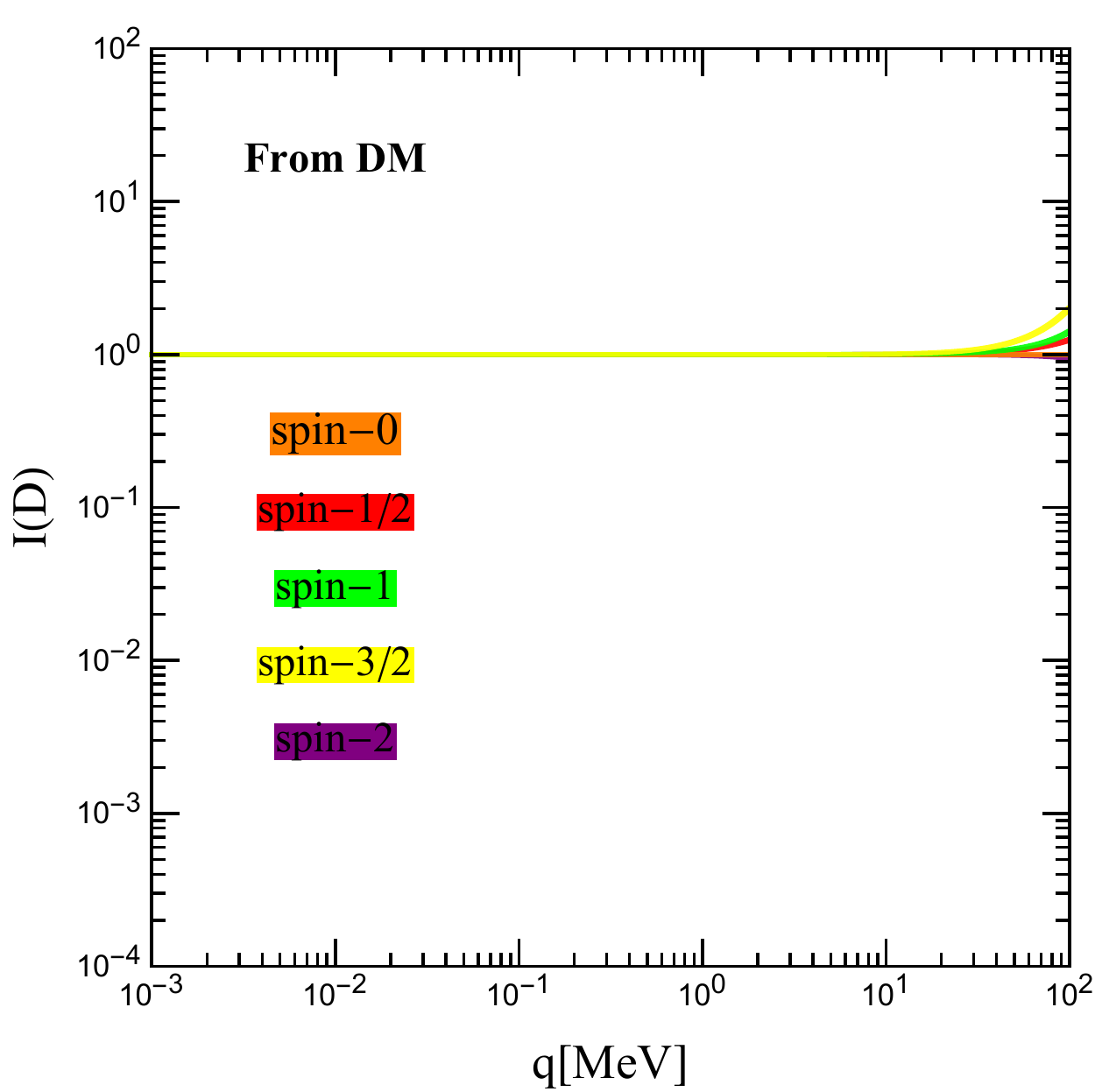}
\caption{The form factor that comes from dark matter interaction is shown this picture. The mass of mediator and dark matter are set as 0.5 MeV and 100 MeV in the calculations, respectively.}	
\label{dm}
\end{figure}

 We can understand this result from the Eq. \eqref{ifdm}, Eq. \eqref{dfdm} and Fig. \ref{dm}.
 In Fig. \ref{dm}, we plot the dependence of $I(D)$ on transferred momentum. It can be seen that all the result of $I(D)$ are a constant 1. So the $I(D)$ from dark matter interaction is momentum-independence. 
 
 Besides, we also find that the form factor will be suppressed by $1/q^{2}$ when mediator mass less than transferred momentum ($q>0.1$) MeV in both SI and SD scattering. In this case, SI and SD scattering all satisfy the long-range interaction. However, for heavy mediator ($q<0.1$ MeV), the form factor of SI scattering will be constant 1 and the form factor of SD scattering will be enhanced by $q^{2}$. In this case, the result of SI scattering satisfies the contact interaction. But, SD scattering is not contact interaction. 
This is because the mass of electron in SD scattering can be canceled by the pseudoscalar with $\gamma_{\star}$ compared to SI scattering from Eq. \eqref{ifdm} and Eq. \eqref{dfdm}. 
\begin{figure}
\centering
\includegraphics[scale=0.60]{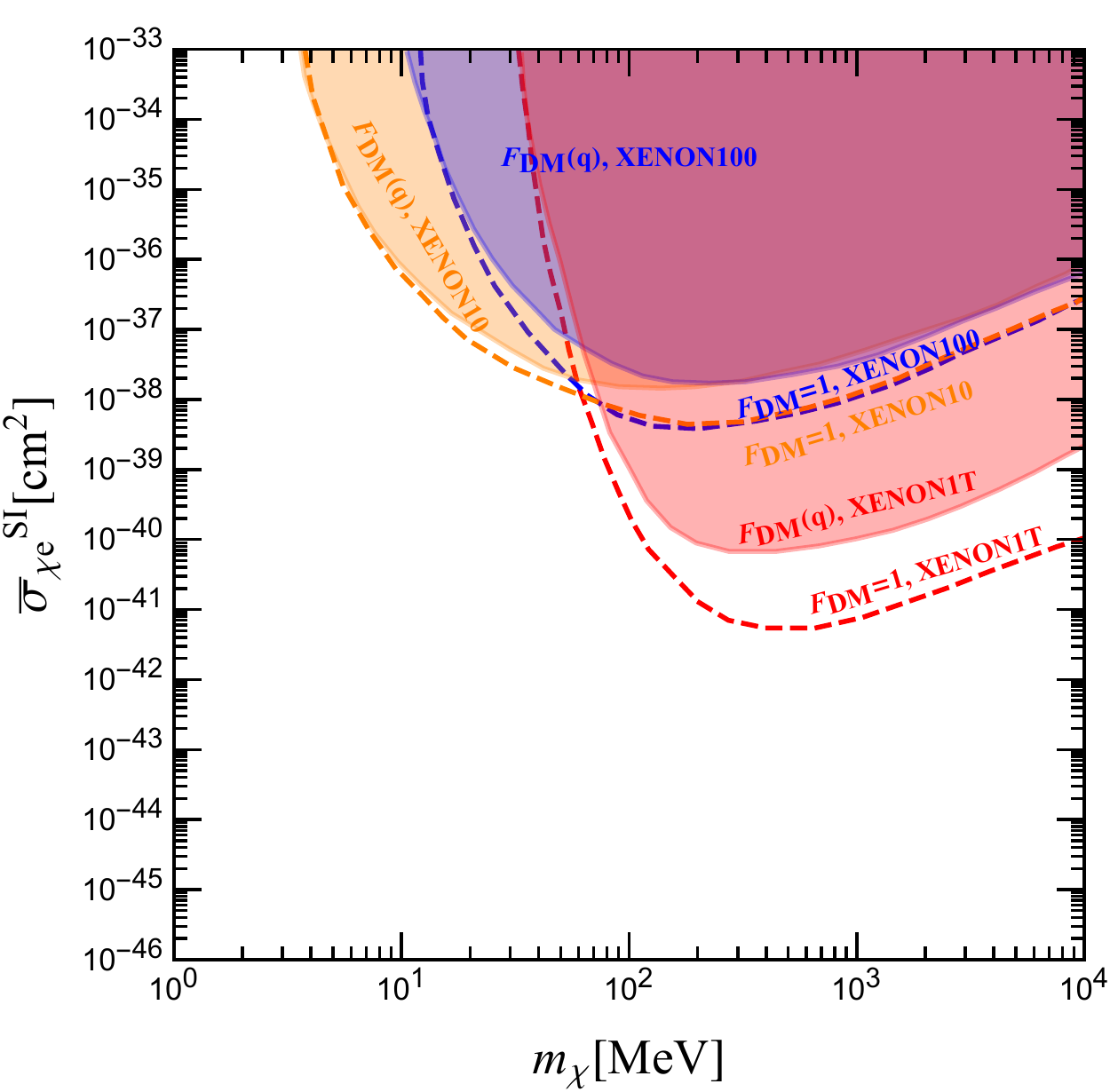}
\includegraphics[scale=0.60]{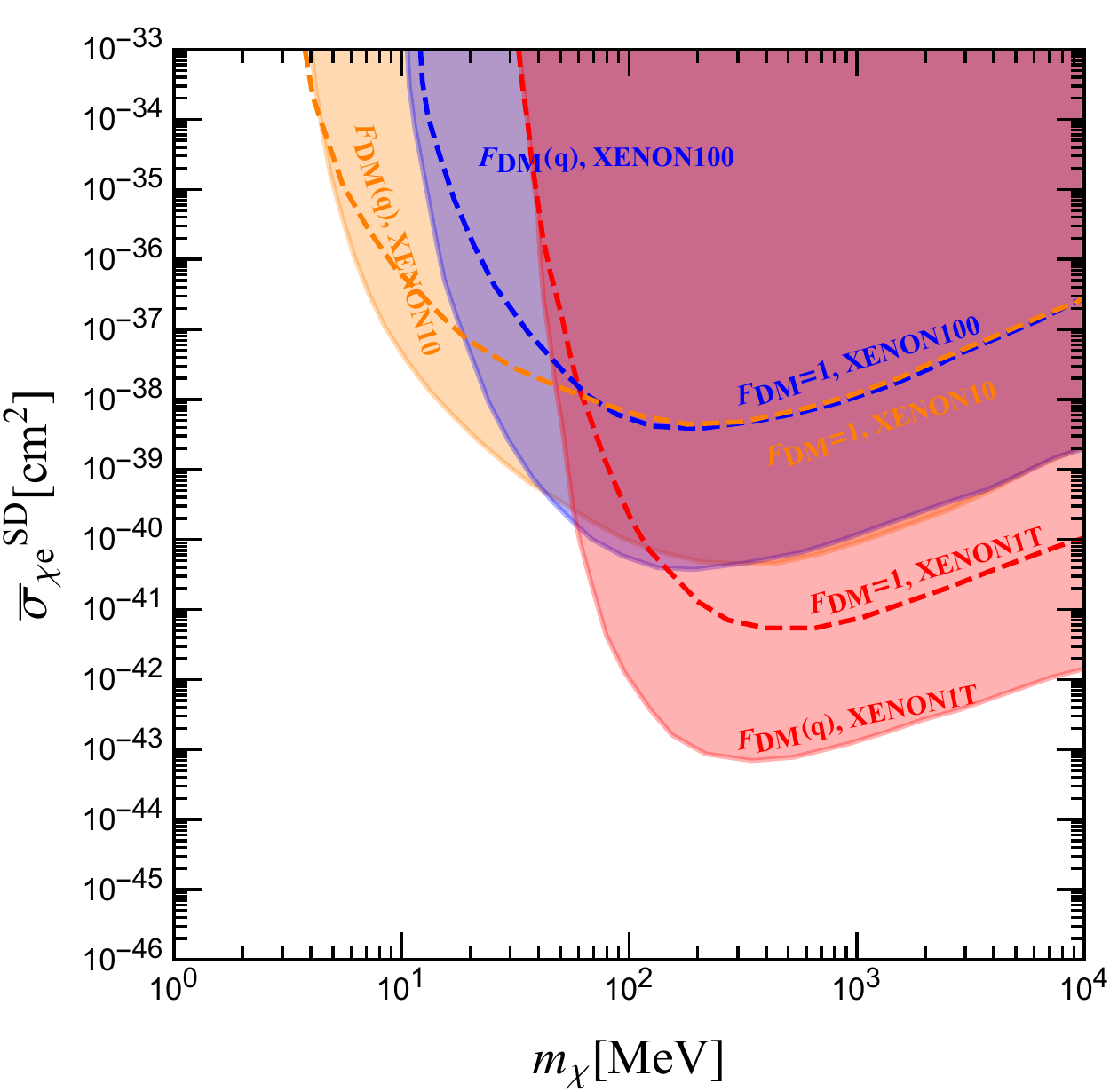}
\caption{ The exclusion limits for the SI (top) and SD (bottom) scattering cross sections from XENON10 (orange), XENON100 (blue) and XENON1T (red) data. The mass of mediator is set as $m_A=0.5$ MeV in the calculations. The result from $F_{\rm DM}=1$ is also shown by dotted lines in each panel.}
\label{cs}
\end{figure}

In order to indicate the effect of SD scattering in cross section with light mediator. Firstly, we plot the numerical result of $F_{\rm DM}=1$ from literature \cite{Essig:2015cda} in Fig. \ref{cs} with dotted lines. As we talked in above, the form factors will have same result at dark matter spin $s=0,~1/2,~1,~3/2,~2$. So the reduced cross section $\bar{\sigma}_{e}$ that dependent on the form factor (this relation can be seen in Eq. \eqref{sigma1}) also will be same when the dark matter is  $0,~1/2,~1,~3/2,~2$. Therefore, we take $s=3/2$ as an example to explore the effect of SD scattering in cross section. The limits of reduced scattering cross section $\bar{\sigma}_{e}^{SI}$(for spin-independent) and $\bar{\sigma}_{e}^{SD}$ (for spin-dependent) at $m_{a}(m_{r})=$ 0.1 MeV is shown in Fig. \ref{cs} with shadow. In which, the red shadow is XENON1T result, blue shadow is XENON100 result and the orange is result from XENON10. From Fig. \ref{cs}, we find that the constrains of SI scattering at light mediator always less than $F_{\rm DM}=1$ (heavy mediator). The reason for this result is that the form factor of spin-independent has suppressive  effect when the transferred momentum greater than 0.1 MeV ($q>0.1$ MeV) in Fig. \ref{fdm}. Especially, in literature \cite{Essig:2015cda}, the constrains of spin-independent scattering cross section from XENON100 can reach $10^{-35}$ $\rm cm^{2}$ (from XENON100) at $\vert F_{\rm DM}\vert^{2}\approx1/q^{2}$. In this case, the mass of mediator is zero. We can see that the effect of suppression from $1/q^{2}$ will be more stronger with smaller mediator mass. The Migdal effect also have similar result in literature \cite{Wang:2021oha}. For SD scattering, the form factor also have suppressive effect at light mediator $(q>0.1$ MeV). However, when $q<0.1$ MeV (heavy mediator), the form factor can be enhanced by $q^{2}$. And the form factor roughly two order of magnitude is improved than $F_{\rm DM}=1$ at $m_{a}(m_{r})=q$. So the bounds of SD scattering cross section at light mediator can increase two order of magnitude than $F_{\rm DM}=1$.



In Fig. \ref{cs}, we also find that the XENON10/100 data have a stronger constraints than XENON1T data in the region of $m_{\text{DM}}<100$ MeV. However, when $m_{\text{DM}}>100$ MeV, the constraints from XENON1T can be larger than XENON10/100 data. This is because that the experiment threshold of XENON10/100 is lower than XENON1T.

  \section{conclusion}\label{sec4}
  The experimental efforts to search for spin-dependent  sub-GeV dark matter motivate consideration of dark matter spin.
  In this paper, we calculated the couplings of dark matter-electron mediated by (pseudo) scalar particles with dark matter spin $s=3/2$ and $s=2$. We also reviewed the result of dark matter spin $s=0$, $s=1/2$ and $s=1$ in literature\cite{Wu:2022jln,Liu:2021avx}. Due to the form factor is model-dependent, we plot form factor from different spin of dark matter in Fig. \ref{fdm} to explore the effect of dark matter spin on interaction. The scattering cross section of SI and SD is plotted in Fig. \ref{cs} to explore the effect of transferring momentum.
  
  Our result show that the form factor will have the same result in any dark matter spin when the mediator is (pseudo)scalar. So the spin of dark matter will not affect the interaction whether SI or SD scattering. The coupling from electron will dominate the interaction.
  Besides, we also find that the form factor $F_{\rm DM}$ will be a constant 1 in SI scattering when the mass of mediator is greater than transferred. However, in SD scattering, the form factor will be enhanced by $q^{2}$ which is caused by the $\gamma_{\star}$ from pseudoscalar. For light mediator, the form factor can be suppressed by the $1/q^{2}$ whether SI or SD scattering. Especially in SI scattering, the mass of mediator is smaller, the suppressive effect in scattering cross section can be more stronger. For SD scattering, the peak value of form factor is much larger than $F_{\rm DM}=1$, therefore the scattering cross section will be enhanced by the transferred momentum compared with $F_{\rm DM}=1$.

  \section{Acknowledge}
  We would like to thank Wenyu Wang and Bin Zhu for many significant guidance on procedural. We also thank Lei Wu for many useful discussions on form factor and SD theory. K-Y W is supported by the Natural Science Foundation of China under grant number 11775012. Acckin is supported by NSFC grant no.11875082.

  \appendix
  
  \section{Conventions and spinor algebra \label{AppenSpinorConven}}
  
  The metric in Minkowski spacetime is chosen as
  \begin{align}
  & g_{\mu\nu}=\text{diag}(+1,-1,-1,-1)
  \end{align}
  For the Levi-Civita symbol, we use the convention $\epsilon_{0123}=1$. In Dirac representation, the $\gamma$-matrices are given by
  \begin{align}
  \label{GammaDirac}
  &\gamma^{0}=\left(\begin{array}{cc}
  I & 0\\
  0 & -I
  \end{array}\right)~,~\gamma^{k}=\left(\begin{array}{cc}
  0 & \sigma_{k}\\
  -\sigma_{k} & 0
  \end{array}\right)
  \end{align}
  It is easy to check that the $\eqref{GammaDirac}$ satisfy the anti-commutation relations
  \begin{align}
  &\{ \gamma^\mu ,  \gamma^\nu \}=2g^{\mu\nu} I
  \end{align}
  The charge conjugation matrix $\mathcal{C}$ needs to satisfy
  \begin{align}
  \label{DefineCp1}
  &\mathcal{C}^{T}=-\mathcal{C}\\
    \label{DefineCp2}
  &(\gamma^{\mu})^{T}=-\mathcal{C}\gamma^{\mu}\mathcal{C}^{-1}
  \end{align}
 Combine the properties $\eqref{DefineCp1}$-$\eqref{DefineCp2}$ with $\eqref{GammaDirac}$, the ansatz of $\mathcal{C}$ is constructed as
 \begin{align}
 &\mathcal{C}=\left(\begin{array}{cccc}
 0 & 0 & 0 & -c\\
 0 & 0 & c & 0\\
 0 & -c & 0 & 0\\
 c & 0 & 0 & 0
 \end{array}\right)
 \end{align}
 in which $c$ is an arbitrary non-zero constant. For convenience sake, one could make $c=1$. In this way, the charge conjugation matrix $\mathcal{C}$ reduces to
 \begin{align}
 &\mathcal{C}=i\gamma^2\gamma^0
 \end{align}
 According to the operator $\mathcal{C}$, the charge conjugation of spinor $\psi$ can be established as
 \begin{align} 
 &\psi^{C}=i\gamma^{0}\mathcal{C}^{-1}\psi^{\star}
 \end{align} 
 In this paper we use the Majorana spinor, namely the $\psi$ should satisfy $\psi=\psi^C$. The Fourier mode expansion of $\psi(x)$ in momentum space is given by
 \begin{align}
 &\psi(x)=\sum_{s=\pm}\int\frac{d^{3}\vec{p}}{2p^{0}(2\pi)^{3}}\big(u_{s}(\vec{p})e^{-ipx}+u_{s}^{C}(\vec{p})e^{ipx}\big)
 \end{align}
 The corresponding solution of the Dirac equation in momentum space is
 \begin{align}
 \label{DiracSpinor}
 &u_s(\vec{p})=\left(\begin{array}{c}
 \sqrt{p^{0}+m}\chi^{(s)}\\
 \sqrt{p^{0}-m}n_{i}\sigma_{i}\chi^{(s)}
 \end{array}\right)
 \end{align}
  in which
  \begin{align}
  \nonumber
  &\chi^{(+1)}=\left(\begin{array}{c}
  e^{-i\phi/2}\cos\theta/2\\
  e^{i\phi/2}\sin\theta/2
  \end{array}\right),\chi^{(-1)}=\left(\begin{array}{c}
  -e^{-i\phi/2}\sin\theta/2\\
  e^{i\phi/2}\cos\theta/2
  \end{array}\right)
  \end{align}
  And we take the four momentum as the following form
  \begin{align}
  \label{4MomentSphere}
  &p^\mu=(p_0,\vec{p})=(\sqrt{\vert\vec{p}\vert^2+m^2},\vert\vec{p}\vert \sin\theta\cos\phi,\vert\vec{p}\vert \sin\theta\sin\phi,\vert\vec{p}\vert \cos\theta)
  \end{align}
  The $\vec{n}$ denotes radial unit vector in spherical polar coordinates
  \begin{align}
  \nonumber
  &\vec{n}=\frac{\vec{p}}{\vert\vec{p}\vert}=(\sin\theta\cos\phi,\sin\theta\sin\phi,\cos\theta)
  \end{align}
  Note that the helicity sum of spinor $\eqref{DiracSpinor}$ satisfies the following formula
  \begin{align}
  \label{HelicitySumSpinor}
  &\sum_{s}u(\vec{p},s)\bar{u}(\vec{p},s)=\gamma^{\mu}p_{\mu}+mI
  \end{align}
  
  Besides, we provide some useful results concerning the Clifford algebra, which could help us compute the trace of multiple $\gamma$-matrices product. One could refer to \cite{Freedman:2012zz} for getting more details. The complete Clifford algebra for spacetime dimension $D$ consists of the following list
  \begin{align}
  \label{BasisOfClifford}
  &\Gamma^A=\{ I, \gamma^\mu, \gamma^{\mu_1 \mu_2}, \gamma^{\mu_1 \mu_2 \mu_3},\dots, \gamma^{\mu_1 \dots \mu_D} \}
  \end{align}
  in which
  \begin{align}
  \nonumber
  &\gamma^{\mu_1 \mu_2}=\frac{1}{2} [\gamma^{\mu_1}, \gamma^{\mu_2}]~,~\gamma^{\mu_1\mu_2\mu_3}=\frac{1}{2} \{ \gamma^{\mu_1}, \gamma^{\mu_2 \mu_3} \}\\
  \nonumber
  &\gamma^{\mu_{1}\mu_{2}\mu_{3}\mu_{4}}=\frac{1}{2}[\gamma^{\mu_{1}},\gamma^{\mu_{2}\mu_{3}\mu_{4}}]~,~\dots\\
  \label{DefineBasisOfGamma}
  &\gamma^{\rho\mu_{1}\dots\mu_{r}}=\frac{1}{2}\gamma^{\rho}\gamma^{\mu_{1}\dots\mu_{r}}+\frac{1}{2}(-1)^{r}\gamma^{\mu_{1}\dots\mu_{r}}\gamma^{\rho}
  \end{align}
  It is easy to check that the indices $\mu_1,~\mu_2,~\dots,~\mu_r$ are totally anti-symmetric in $\gamma^{\mu_{1}\mu_{2}\dots\mu_{r}}$. An important characteristic of $\gamma^{\mu_{1}\mu_{2}\dots\mu_{r}}~(D > r\geq 1)$ is that all these matrices are traceless. For the highest rank matrix with $r=D$, it is traceless only for even spacetiem dimension $D$. In order to decompose the product of the multiple Gamma matrices into the basis given by $\eqref{BasisOfClifford}$, we provide some useful tricks
  \begin{align}
  \nonumber
  &\gamma^{\mu_{1}\mu_{2}\dots\mu_{r}}\gamma^{\rho}=2\gamma^{\mu_{1}\mu_{2}\dots\mu_{r-1}}g^{\mu_{r}\rho}-2\gamma^{\mu_{1}\mu_{2}\dots\mu_{r-2}\mu_{r}}g^{\mu_{r-1}\rho}\\
  \nonumber
  &\quad\quad\quad\quad\quad\quad+\dots+2(-1)^{r-1}\gamma^{\mu_{2}\dots\mu_{r}}g^{\mu_{1}\rho}\\
   \label{trickCliff1}
  &\quad\quad\quad\quad\quad\quad+(-1)^{r}\gamma^{\rho}\gamma^{\mu_{1}\mu_{2}\dots\mu_{r}}	
  \end{align}
  Combine with $\eqref{DefineBasisOfGamma}$, the $\eqref{trickCliff1}$ is rewritten as
  \begin{align}
  \nonumber
  &\hspace{-7mm}\gamma^{\mu_{1}\dots\mu_{r}}\gamma^{\rho}=(-1)^{r}\gamma^{\rho\mu_{1}\dots\mu_{r}}+\gamma^{\mu_{1}\dots\mu_{r-1}}g^{\mu_{r}\rho}-\gamma^{\mu_{1}\dots\mu_{r-2}\mu_{r}}g^{\mu_{r-1}\rho}\\
\label{trickCliff2}
  &\quad\quad~+\dots+(-1)^{r-1}\gamma^{\mu_{2}\dots\mu_{r}}g^{\mu_{1}\rho}
  \end{align}
  In case of even dimension $D=2m$, the highest rank Clifford algebra element is defined as
  \begin{align}
  &\gamma_{\star}=(-i)^{m+1}\gamma_{0}\gamma_{1}\dots\gamma_{D-1}
  \end{align}
  which satisfies
  \begin{align}
   \label{ProperGamStar1}
  &\gamma_{\star}^{2}=I~,~\gamma_{\star}=\gamma_{\star}^{\dagger}
  \end{align}
  Meanwhile, the $\gamma_{\star}$ commutes with all even-rank elements in $\eqref{BasisOfClifford}$ and anti-commutes with all odd rank elements, namely
  \begin{align}
  \label{ProperGamStar2}
  &\{\gamma_{\star},\gamma^{\mu_{0}\dots\mu_{2k}}\}=0~,~[\gamma_{\star},\gamma^{\mu_{0}\dots\mu_{2k+1}}]=0
  \end{align}
  From the above properties and equation $\eqref{trickCliff1}$, one could derive the duality relations between rank $r$ and rank $D-r$ sectors in $\eqref{BasisOfClifford}$
  \begin{align}
  \label{dualRelaEven}
  &\hspace{-4.5mm}\gamma^{\nu_{1}\dots\nu_{r}}\gamma_{\star}=\frac{(i)^{m+1}(-1)^{r+1}}{(D-r)!}\epsilon^{\nu_{1}\dots\nu_{r}\mu_{r+1}\dots\mu_{D}}\gamma_{\mu_{D}\dots\mu_{r+1}}
  \end{align}
  It is necessary to stress again that the $\eqref{dualRelaEven}$ is only valid for even spacetime dimension, i.e. $D=2m$.
  
  In this paper, we restrict attentions to case of $D=4$. And some typical examples concerning trace of multiple $\gamma$-matrices product are considered, which are useful in calculating the square of the scattering amplitude which include the contributions from fermions. The simplest form is
  \begin{align}
  \label{DoubleGamma}
  &\gamma_{\mu}\gamma_{\nu}=g_{\mu\nu}I+\gamma_{\mu\nu}
  \end{align}
  It is easy to obtain
  \begin{align}
  \label{TraceProdSimGam}
  &\text{Tr}(\gamma_{\mu}\gamma_{\nu})=4g_{\mu\nu}
  \end{align}
  And then, we consider a more complicated form
  \begin{align}
  \nonumber
  &\gamma^{\alpha\beta}\gamma^{\lambda}\gamma^{\rho}=-\gamma^{\rho\alpha\beta\lambda}+\gamma^{\alpha\beta}g^{\lambda\rho}+2\gamma^{\lambda[\alpha}g^{\beta]\rho}\\
  &\quad\quad\quad~~~-2\gamma^{\rho[\alpha}g^{\beta]\lambda}+2g^{\rho[\alpha}g^{\beta]\lambda}I
  \end{align}
  from which one could obtain
  \begin{align}
  \label{DecomDouGamma}
  &\gamma_{\alpha\beta}\gamma^{\lambda\rho}=\gamma_{~~\alpha\beta}^{\lambda\rho}+4\delta_{[\alpha}^{[\lambda}\gamma_{~\beta]}^{\rho]}+2\delta_{[\alpha}^{[\rho}\delta_{\beta]}^{\lambda]}I
  \end{align}
  And hence the trace of $\eqref{DecomDouGamma}$ has the expression
  \begin{align}
  &\text{Tr}(\gamma^{\mu\nu}\gamma^{\lambda\rho})=4g^{\mu[\rho}g^{\lambda]\nu}-4g^{\nu[\rho}g^{\lambda]\mu}
  \end{align}
  Finally, due to $\eqref{trickCliff2}$ and $\eqref{dualRelaEven}$, we can decompose
  \begin{align}
  \nonumber
  &\gamma^{\alpha\beta}\gamma^{\lambda}=\gamma^{\alpha\beta\lambda}+\gamma^{\alpha}g^{\beta\lambda}-\gamma^{\beta}g^{\alpha\lambda}\\
  \label{ThreeGamma}
  &\quad\quad~~=-i\epsilon^{\alpha\beta\lambda\mu}\gamma_{\mu}\gamma_{\star}+2\gamma^{[\alpha}g^{\beta]\lambda}
  \end{align}
  According to $\eqref{ProperGamStar1}, \eqref{ProperGamStar2}$ and $\eqref{TraceProdSimGam}$, it allows us to get
  \begin{align}
  \nonumber
  &\quad\text{Tr}(\gamma^{\alpha_{1}\beta_{1}}\gamma^{\lambda_{1}}\gamma^{\alpha_{2}\beta_{2}}\gamma^{\lambda_{2}})\\
  &=-4!g^{\alpha_{1}[\underline{\alpha_{2}}}g^{\beta_{1}\underline{\beta_{2}}}g^{\lambda_{1}\underline{\lambda_{2}}]}+16g^{\lambda_{2}[\beta_{2}}g^{\alpha_{2}][\alpha_{1}}g^{\beta_{1}]\lambda_{1}}
  \end{align}
  where we have used the identity $\epsilon^{\mu\alpha_{1}\beta_{1}\lambda_{1}}\epsilon_{\mu}^{~\alpha_{2}\beta_{2}\lambda_{2}}=-3!g^{\alpha_{1}[\underline{\alpha_{2}}}g^{\beta_{1}\underline{\beta_{2}}}g^{\lambda_{1}\underline{\lambda_{2}}]}$. Meanwhile, it is necessary to indicate that the indices with underline are totally anti-symmetric.
  
  \section{Wave function and the helicity sum for free massive gravitino \label{IntroToGravitino}}
  
  The massive gravitino is described by the Rarita-Schwinger Lagrangian
  \begin{align}
  \label{MassiveGravitinoAction}
  \mathcal{L}=-\frac{1}{2}\epsilon^{\mu\nu\rho\sigma}\bar{\Psi}_{\mu}\gamma_{\star}\gamma_{\nu}\partial_{\rho}\Psi_{\sigma}-\frac{1}{2}m_{3/2}\bar{\Psi}_{\mu}\gamma^{\mu\nu}\Psi_{\nu}
  \end{align}
  From $\eqref{MassiveGravitinoAction}$, the corresponding equation of motion is obtained as
  \begin{align}
  \label{EOMsGravitinoV1}
  &\epsilon^{\mu\nu\rho\sigma}\gamma_{\star}\gamma_{\nu}\partial_{\rho}\Psi_{\sigma}+m_{3/2}\gamma^{\mu\nu}\Psi_{\nu}=0
  \end{align}
  First, we multiply equation $\eqref{EOMsGravitinoV1}$ by $\partial_\mu$, it yields
  \begin{align}
  \label{EOMsGravitinoModifiV1}
  &m_{3/2}\big(\gamma^{\mu}\gamma^{\nu}\partial_{\mu}\Psi_{\nu}-\gamma^{\nu}\gamma^{\mu}\partial_{\mu}\Psi_{\nu}\big)=0
  \end{align}
  Next, by operating a $\gamma_\mu$ on $\eqref{EOMsGravitinoV1}$, we give
  \begin{align}
  \label{EOMsGravitinoModifiV2}
  &i(/\!\!\!\partial\gamma^{\sigma}\Psi_{\sigma}-\gamma^{\sigma}/\!\!\!\partial\Psi_{\sigma})+3m_{3/2}\gamma^{\nu}\Psi_{\nu}=0
  \end{align}
  Note that we have used the duality relation $\eqref{dualRelaEven}$ in deriving the equation $\eqref{EOMsGravitinoModifiV2}$. If one chooses the gauge
  \begin{align}
  \label{GaugeConstrain}
  &\gamma^\mu \Psi_\mu=0
  \end{align}
  And then the equations $\eqref{EOMsGravitinoModifiV1}$-$\eqref{EOMsGravitinoModifiV2}$ simultaneously imply
  \begin{align}
  \label{ReduceEqua}
  &\partial^{\nu}\Psi_{\nu}=0
  \end{align}
  By utilizing the relations $\eqref{DoubleGamma}$ and $\eqref{ThreeGamma}$, we can rewrite the $\eqref{MassiveGravitinoAction}$ as
  \begin{align}
  \label{MassiveGravitinoActionV1}
  &\hspace{-10.4mm}i\big(2\eta^{\sigma[\mu}\gamma^{\rho]}+\gamma^{\mu\rho}\gamma^{\sigma}\big)\partial_{\rho}\Psi_{\sigma}+m_{3/2}\big(\gamma^{\mu}\gamma^{\nu}\Psi_{\nu}-\eta^{\mu\nu}\Psi_{\nu}\big)=0
  \end{align}
  Together with results $\eqref{GaugeConstrain}$-$\eqref{ReduceEqua}$, the $\eqref{MassiveGravitinoActionV1}$ can be further simplified to
  \begin{align}
\label{EOMsGravitinoV2}
  &i\gamma^{\rho}\partial_{\rho}\Psi^{\mu}-m_{3/2}\Psi^{\mu}=0
  \end{align}
  Besides, since the $\Psi_\mu(x)$ is Majorana spinor, it obeys $\Psi_\mu(x)=\Psi^C_\mu(x)$. In momentum space, the $\Psi_\mu (x)$ can be expanded by the following mode function
  \begin{align}
  \label{ModeExpanGraviti}
  &\hspace{-8.2mm}\Psi_{\mu}(x)=\sum_{\lambda}\int\frac{d^{3}\vec{p}}{2p^{0}(2\pi)^{3}}\big(\tilde{\Psi}_{\mu}^{(\lambda)}(\vec{p})e^{-ipx}+\tilde{\Psi}_{\mu}^{(\lambda)C}(\vec{p})e^{ipx}\big)
  \end{align}
  Plugging the mode expansion $\eqref{ModeExpanGraviti}$ into the equations $\eqref{GaugeConstrain}$, $\eqref{ReduceEqua}$ and $\eqref{EOMsGravitinoV2}$, it yields
  \begin{align}
    \label{GaugeConstrainP}
  &\gamma^{\mu}\tilde{\Psi}_{\mu}^{(\lambda)}(\vec{p})=0\\
  \label{ReduceEquaP}
  &p^{\mu}\tilde{\Psi}_{\mu}^{(\lambda)}(\vec{p})=0\\
  \label{EOMsGravitinoV2P}
  &(\gamma^{\nu}p_{\nu}-m_{3/2}I)\tilde{\Psi}_{\mu}^{(\lambda)}(\vec{p})=0		
  \end{align}
  The wave function $\tilde{\psi}_\mu$ can be constituted by the wave function $u$ for massive spin-$\frac{1}{2}$ field and the polarization vector $\epsilon_\mu$ for massive spin-$1$ field \cite{Auvil:1966eao}
  \begin{align}
  \label{GravitinoPsi}
  &\tilde{\psi}_{\mu}^{(\lambda)}(\vec{p})=\sum_{m,s}\delta_{\lambda,m+s}C_{\lambda;m,s}^{\frac{3}{2};1,\frac{1}{2}}u_{s}(\vec{p})\epsilon_{\mu}^{(m)}(\vec{p})
  \end{align}
  in which the $C_{\lambda;m,s}^{\frac{3}{2};1,\frac{1}{2}}$ is the Clebsch-Gordan coefficients in term of the addition of spin -$\frac{1}{2}$ particle with spin-$1$ particle, whose value is listed in Table.\ref{CGHalfwithOne}.
  \begin{table}[!ht]
  	\begin{center}
  		\begin{tabular}{|c|c|c|c|}
  			\hline
  			~ & $m=-1$ & $m=0$ & $m=+1$ \\
  			\hline
  			$s=-\frac{1}{2}$ & $1$ & $\sqrt{\frac{2}{3}}$ & $\sqrt{\frac{1}{3}}$ \\
  			\hline
  			$s=+\frac{1}{2}$ & $\sqrt{\frac{1}{3}}$ & $\sqrt{\frac{2}{3}}$ & $1$\\
  			\hline
  		\end{tabular}
  		\caption{This table shows Clebsch-Gordan coefficients in term of the addition of spin -$\frac{1}{2}$ particle with spin-$1$ particle.}
  		\label{CGHalfwithOne}
  	\end{center}
  \end{table}
For a massive spin-$1$ boson, the polarization vectors quantized with respect to $4$-momentum $\eqref{4MomentSphere}$ can be defined as
  \begin{align}
  \label{Spin1Com0}
 &\epsilon_{\mu}^{(0)}(\vec{p})=-\frac{1}{m}\left(\begin{array}{c}
  -\vert\vec{p}\vert\\
  E\sin\theta\cos\phi\\
  E\sin\theta\sin\phi\\
  E\cos\theta
  \end{array}\right)\\
  \label{Spin1PlusMinus}
&\epsilon_{\mu}^{(\pm1)}(\vec{p})=\frac{1}{\sqrt{2}}\left(\begin{array}{c}
0\\
\pm\cos\theta\cos\phi-i\sin\phi\\
\pm\cos\theta\sin\phi+i\cos\phi\\
\mp\sin\theta
\end{array}\right)		  
\end{align}
which obey the orthogonal and normalized relations
\begin{align}
&g^{\mu\nu}~\epsilon_{\mu}^{(m)}(\vec{p})~\epsilon_{\nu}^{(m^{\prime})}(\vec{p})=-\delta^{mm^{\prime}}
\end{align}  
Besides, the polarization vectors $\eqref{Spin1Com0}$-$\eqref{Spin1PlusMinus}$ also satisfy the following conditions
\begin{align}
\label{PolarOrthogonal}
&p^\mu \epsilon^{(m)}_\mu (\vec{p})=p^\mu \big(\epsilon^{(m)}_\mu (\vec{p})\big)^\star=0\\
\label{PolarSumVector}
&\sum_{m}\epsilon_{\mu}^{(m)}(\vec{p})\epsilon_{\nu}^{(m)}(\vec{p})=-g_{\mu\nu}+\frac{P_{\mu}P_{\nu}}{m^{2}}
\end{align}
By using the composite structure $\eqref{GravitinoPsi}$ and the helicity sum formulas $\eqref{HelicitySumSpinor}$,$\eqref{PolarSumVector}$, one could find
\begin{align}
\nonumber
&\mathcal{P}_{\mu\nu}=\sum_{\lambda}\tilde{\psi}_{\mu}^{(\lambda)}(\vec{p})\tilde{\psi}_{\nu}^{(\lambda)\dagger}(\vec{p})=-(/\!\!\!p-m_{3/2})\bigg((g_{\mu\nu}-\frac{p_{\mu}p_{\nu}}{m_{3/2}^{2}})\\
\label{HeliciSumGravitino}
&\quad\quad-\frac{1}{3}(g_{\mu\rho}-\frac{p_{\mu}p_{\rho}}{m_{3/2}^{2}})(g_{\nu\beta}-\frac{p_{\nu}p_{\beta}}{m_{3/2}^{2}})\gamma^{\rho}\gamma^{\beta}\bigg)
\end{align}
It is straightforward to examine that the following relations hold for $\eqref{HeliciSumGravitino}$
\begin{align}
\nonumber
&\gamma^{\mu}\mathcal{P}_{\mu\nu}(\vec{p})=\mathcal{P}_{\mu\nu}(\vec{p})\gamma^{\nu}=0\\
\nonumber
&p^{\mu}\mathcal{P}_{\mu\nu}(\vec{p})=\mathcal{P}_{\mu\nu}(\vec{p})p^{\nu}=0\\
\nonumber
&(\gamma^{\nu}p_{\nu}-m_{3/2}I)\mathcal{P}_{\mu\nu}(\vec{p})=\mathcal{P}_{\mu\nu}(\vec{p})(\gamma^{\nu}p_{\nu}-m_{3/2}I)=0
\end{align}
which are consistent with the equations $\eqref{GaugeConstrainP}$-$\eqref{EOMsGravitinoV2P}$.

\section{A phenomenological Lagrangian about massive KK-gravitons derive from warped extra-dimensions \label{EffActionFromBrane}}
  
   In this part, our purpose is to construct a phenomenological Lagrangian including the massive KK-gravitons and radions from a simplified version of the Randall-Sundrum model \cite{Randall:1999ee}. At the beginning, we need to introduce some notations. The coordinates of the bulk spacetime is denoted by $X^M$, while the corresponding metric is $G_{MN}(X)$. We use $\tilde{x}^{\tilde{\mu}}$ to represent the intrinsic coordinate of the 3-brane, and the corresponding induced metric is $g_{\tilde{\mu}\tilde{\nu}}(\tilde{x})$. Besides, the $local~Lorentz$ coordinates on brane are characterized by $x^\mu$, for which the line element is described by the Minkowski metric $\eta_{\mu\nu}$. Note that $g_{\tilde{\mu}\tilde{\nu}}(\tilde{x})$ and $\eta_{\alpha\beta}$ can be transfered into each other through the vielbein $e^\alpha_{~\tilde{\mu}}$, namely
 \begin{align}
 &e^\alpha_{~\tilde{\mu}}(\tilde{x}) \eta_{\alpha\beta} e^\beta_{~\tilde{\nu}}(\tilde{x})=g_{\tilde{\mu}\tilde{\nu}}(\tilde{x})\\
 &e_{\alpha}^{~\tilde{\mu}}(\tilde{x})g_{\tilde{\mu}\tilde{\nu}}(\tilde{x})e_{\beta}^{~\tilde{\nu}}(\tilde{x})=\eta_{\alpha\beta}
 \end{align}
 At the same time, the vielbein also satisfy the following orthogonal relations
 \begin{align}
 &e_{~\tilde{\mu}}^{\alpha}e_{\alpha}^{~\tilde{\nu}}=\delta_{\tilde{\mu}}^{\tilde{\nu}}~,~e_{~\tilde{\mu}}^{\alpha}e_{\beta}^{~\tilde{\mu}}=\delta_{\beta}^{\alpha}
 \end{align}
 
 This warped extra-dimensional model is described by the following action
 \begin{align}
 \label{BraneBulkSystem}
&S=S_{bulk}+S_{UV}+S_{IR}\\
\nonumber
&S_{bulk}=\frac{1}{2\kappa^{3}}\int d^{4}\tilde{x}\int_{-\pi}^{+\pi}d\varphi\sqrt{G}(\mathcal{R}-2\Lambda_{\text{bulk}})\\
\nonumber
&S_{\text{UV}}=\int d^{4}\tilde{x}\int_{-\pi}^{+\pi}d\varphi\sqrt{-g^{\text{UV}}}(\mathcal{L}_{\text{UV}}-\Lambda_{\text{UV}})\delta(\varphi)\\
\nonumber
&S_{\text{IR}}=\int d^{4}\tilde{x}\int_{-\pi}^{+\pi}d\varphi\sqrt{-g^{\text{IR}}}(\mathcal{L}_{\text{IR}}-\Lambda_{\text{IR}})\delta(\varphi-\pi)
 \end{align}
where $(g^{\text{UV}})_{\tilde{\mu} \tilde{\nu}}$ and $(g^{\text{IR}})_{\tilde{\mu} \tilde{\nu}}$ represent the induced metric of ultraviolet brane and infrared brane localized on the $\varphi=0$ and $\varphi=\pi$ respectively, while the parameter $\kappa$ is $\frac{1}{2\kappa^{2}}=\frac{1}{16\pi G_{\text{5}}}=\frac{M_{\text{5}}^{3}}{2}$. Furthermore, we use $\Lambda_{\text{UV}}$ and $\Lambda_{\text{IR}}$ to represent the vacuum energy terms on the brane. Note that there only exists the gravitational interaction in bulk spacetime, and the SM particles are localized on branes. If the contributions of matter field are neglected, i.e. only the vacuum energy exist in the bulk and on the branes, the corresponding Einstein equations are solved by
\begin{align}
\nonumber
&ds^{2}\!=\!G_{MN}dX^{M}dX^{N}\!=g_{\tilde{\mu}\tilde{\nu}}d\tilde{x}^{\tilde{\mu}}d\tilde{x}^{\tilde{\nu}}-r_{c}^{2}d\varphi^{2}\\
\label{VacumBulkBraneSol}
&\quad~\!=\!\text{e}^{-2kr_c\vert \varphi\vert}\eta_{\mu\nu}dx^{\mu}dx^{\nu}\!-\!r_{c}^{2}d\varphi^{2}
\end{align}
 in which the $\varphi\in[-\pi,\pi]$ denotes an extra compactified dimension of space with an $S^1/Z_2$ orbifold symmetry, while the size of the extra dimension is characterized by $r_c$. Meanwhile, the warping parameter $k$ is defined as $k=\sqrt{-\Lambda_{\text{bulk}}/6}$. It is necessary to emphasize that the relation $\Lambda_{\text{UV}}=-\Lambda_{\text{IR}}=6M_{\text{5}}k$ should hold in order to preserve the 4-dimensional Poincare invariance exhibited in $\eqref{VacumBulkBraneSol}$. For brevity, we introduce the abbreviation $A(\varphi)=\text{e}^{-kr_{c}\vert\varphi\vert}$. If the particles are generated from the quantum fluctuations, the metric is taken to be perturbed around the vacuum solution, namely
 \begin{align}
 \label{PerMetric}
 &\hspace{-1.2mm}ds^{2}\!=\!A(z)^{2}\big(\!\text{e}^{-2u}(\eta_{\mu\nu}\!\!+\!\!\kappa \hat{h}_{\mu\nu})dx^{\mu}dx^{\nu}\!\!-\!\!(1+2\hat{u})^{2}dz^{2}\big)
 \end{align}
 where the new coordinate $z$ is related to $\varphi$ through
 \begin{align}
 &dz=r_{c}A(\varphi)^{-1}d\varphi~,~\frac{\partial}{\partial z}=r_{c}^{-1}A(\varphi)\frac{\partial}{\partial\varphi}     
 \end{align}
 And the symmetric tensor field $h_{\mu\nu}(x,z)$ corresponds to the spin-2 graviton in the bulk spacetime, while the scalar perturbation in the 5th dimension is denoted by $\hat{u}(x,z)$.
 In fact, the component of vector perturbation in $\eqref{PerMetric}$ has been removed by choosing the appropriate gauge \cite{Callin:2004zm}. By employing the Kaluza-Klein (KK) decomposition, the dependence of the fields in extra-dimensional coordinate is factorized into
 \begin{align}
 \label{GravitonExcita}
 &\hat{h}_{\mu\nu}(x,z)=\sum_{n=0}^{\infty}\frac{1}{\sqrt{r_{c}}}h_{\mu\nu}^{(n)}(x)\chi_{n}(\varphi(z))\\
 &\hat{u}(x,z)=\kappa\frac{\hat{r}(x)}{2\sqrt{6}}A(z)^{-2},\hat{r}(x)=\frac{1}{\sqrt{r_{c}}}\psi_{r}\boldsymbol{{r}}(x)		
 \end{align}
 where the $h_{\mu\nu}^{(n)}(x)$ and $\hat{r}(x)$ represent the KK-gravitons and radion localized on the brane with fixed $z$. After substituting the ansatz $\eqref{GravitonExcita}$ into the Einstein equation, we obtain a wave equation describing the free massive gravitons and the following differential equation restricting the $\psi_n(z)$ \cite{Davoudiasl:1999jd}
 \begin{align}
&\frac{d}{dz}\big(A(z)^{3}\frac{d\chi_{n}(z)}{dz}\big)=-m_{n}^{2}A(z)^{3}\chi_{n}(z)
 \end{align}
 in which $m_n$ is the mass of $h_{\mu\nu}^{(n)}$, while the orthogonal and normalizable relation for $\chi_n$ is found to be
 \begin{align}
&\int_{z(\varphi=-\pi)}^{z(\varphi=+\pi)}dz~\frac{A(z)^{3}}{r_{c}}\chi_{n}(z)\chi_{m}(z)=\delta_{n,m}     
 \end{align}
 For the sake of simplifying the problem considerably, we will neglect the effects of KK tower (Hereaftr, the index $n$ in $h^{(n)}_{\mu\nu}$ will be thrown away). Although this simplification is coarse more or less, it has a very small impact on magnitude of the scattering amplitude which includes the KK-gravitons. Especially, in this paper, we primarily focus on the form factor which is proportional to the ratio of scattering amplitude with different transferred momentum. And hence the effects of KK tower on the form factor is so tiny that we can ignore it narually.
 
 We assume the standard model particles are localized on the $UV$ brane. In this work, we mainly focus on the scattering between the KK-graviton dark matter and electron. And hence, only the localization of fermionic matter is considered in $\mathcal{L}_{UV}$, namely
 \begin{align}
 &\hspace{-1mm}\mathcal{L}_{\text{UV}}\!=\!\sqrt{-g^{\text{UV}}}\big\{\!\text{i}\bar{\psi}e_{\rho}^{~\tilde{\mu}}\gamma^{\rho}D_{\tilde{\mu}}\psi\!-\!\!m_{e}\bar{\psi}\psi\!\big\}\\
 \nonumber
 &D_{\tilde{\mu}}=\partial_{\tilde{\mu}}\!+\!\!\frac{1}{4}\omega_{\tilde{\mu}}^{~\alpha\beta}\gamma_{\alpha\beta}
 \end{align}
 where the $\omega_{\tilde{\mu}}^{~\alpha\beta}$ is the spin-connection defined as
\begin{align}
\label{SpinConnec}
&\omega_{\tilde{\mu}[\alpha\beta]}=\eta_{[\alpha\lambda}\Gamma_{\tilde{\mu}\tilde{\nu}}^{\tilde{\rho}}e_{~\tilde{\rho}}^{\lambda}e_{\beta]}^{~\tilde{\nu}}-\eta_{[\alpha\lambda}e_{\beta]}^{~\tilde{\nu}}\cdot\partial_{\tilde{\mu}}e_{~\tilde{\nu}}^{\lambda}
\end{align}
Notice that the existence of $\omega_{\tilde{\mu}}^{~\alpha\beta}$ preserves that $D_{\tilde{\mu}}\psi$ behaves as a spinor under the local Lorentz transformations, namely
\begin{align}
\nonumber &\psi(\tilde{x})\!\!\to\!\!\text{e}^{-\frac{1}{4}\lambda^{\alpha\beta}(\tilde{x})\gamma_{\alpha\beta}}\!\psi(\tilde{x})\Longrightarrow D_{\tilde{\mu}}\psi(\tilde{x})\!\!\to\!\!\text{e}^{-\frac{1}{4}\lambda^{\alpha\beta}(\tilde{x})\gamma_{\alpha\beta}}\!D_{\tilde{\mu}}\psi(\tilde{x})
\end{align}
From \eqref{BraneBulkSystem}, we see that the UV brane is localized on $\varphi=0$, and the corresponding induced metric (perturbation at $O(\kappa)$ order) reads
\begin{align}
\label{InduceBraneUV}
&ds_{\text{UV}}^{2}=\big(\eta_{\tilde{\mu}\tilde{\nu}}+\frac{\kappa}{\sqrt{r_{c}}}h_{\tilde{\mu}\tilde{\nu}}-\frac{\kappa}{\sqrt{6r_{c}}}\eta_{\tilde{\mu}\tilde{\nu}}\boldsymbol{r}\big)d\tilde{x}^{\mu}d\tilde{x}^{\nu}
\end{align}
In term of \eqref{InduceBraneUV}, the vielbein can be constructed as
\begin{align}
\label{VeilbeinV1}
&e_{~\tilde{\mu}}^{\alpha}=\delta_{~\tilde{\mu}}^{\alpha}+\frac{\kappa}{2\sqrt{r_{c}}}\delta_{~\tilde{\nu}}^{\alpha}\hat{h}_{~\tilde{\mu}}^{\tilde{\nu}}-\frac{\kappa}{2\sqrt{6r_{c}}}\boldsymbol{r}\delta_{~\tilde{\mu}}^{\alpha}\\
\label{VeilbeinV2}
&e_{\alpha}^{~\tilde{\mu}}=\delta_{\alpha}^{~\tilde{\mu}}-\frac{\kappa}{2\sqrt{r_{c}}}\delta_{\alpha}^{~\tilde{\nu}}\hat{h}_{\tilde{\nu}}^{~\tilde{\mu}}+\frac{\kappa}{2\sqrt{6r_{c}}}\boldsymbol{r}\delta_{\alpha}^{~\tilde{\mu}}
\end{align}
Notably, the $\eqref{VeilbeinV1}$ and $\eqref{VeilbeinV2}$ are valid only at the first order of $\kappa$. Actually, in case of the UV brane \eqref{InduceBraneUV}, the coordinates $\tilde{x}_\mu$ is same to the Minkowski one. And it is worthwhile to indicate that the basic information of spacetime perturbation could also be encoded in vielbein $\eqref{VeilbeinV1}$-$\eqref{VeilbeinV2}$ which play a crucial role in constructing the perturbative expansion of fermionic action. From $\eqref{InduceBraneUV}$, the perturbative connection $\Gamma_{\tilde{\mu}\tilde{\nu}}^{\tilde{\rho}}$ has the expression
\begin{align}
\nonumber
&\Gamma_{\tilde{\mu}\tilde{\nu}}^{\tilde{\rho}}=\frac{\kappa}{2\sqrt{r_{c}}}\eta^{\tilde{\rho}\tilde{\sigma}}\big(\partial_{\tilde{\mu}}h_{\tilde{\nu}\tilde{\sigma}}+\partial_{\tilde{\nu}}h_{\tilde{\sigma}\tilde{\mu}}-\partial_{\tilde{\sigma}}h_{\tilde{\mu}\tilde{\nu}}\big)\\
\label{PerExpanConnec}
&\quad\quad~~-\frac{\kappa}{2\sqrt{6r_{c}}}\big(\delta_{\tilde{\nu}}^{\tilde{\rho}}\partial_{\tilde{\mu}}\boldsymbol{r}+\delta_{\tilde{\mu}}^{\tilde{\rho}}\partial_{\tilde{\nu}}\boldsymbol{r}-\eta^{\tilde{\rho}\tilde{\sigma}}\eta_{\tilde{\mu}\tilde{\nu}}\partial_{\tilde{\sigma}}\boldsymbol{r}\big)
\end{align}
By combining $\eqref{SpinConnec}$ with $\eqref{VeilbeinV1}$-$\eqref{PerExpanConnec}$, the $e_{\lambda}^{~\tilde{\mu}}\omega_{\tilde{\mu}[\alpha\beta]}$ is expanded to
\begin{align}
\nonumber
&e_{\lambda}^{~\tilde{\mu}}\omega_{\tilde{\mu}[\alpha\beta]}=\frac{\kappa}{2}\big(\frac{1}{\sqrt{6}}\eta_{\beta\lambda}\partial_{\alpha}\boldsymbol{r}(x)-\frac{1}{\sqrt{6}}\eta_{\lambda\alpha}\partial_{\beta}\boldsymbol{r}(x)\\
\label{VeilbinSpinConnec}
&\quad\quad\quad\quad+\partial_{\beta}h_{\alpha\lambda}(x)-\partial_{\alpha}h_{\lambda\beta}(x)\big)    
\end{align}
Meanwhile, from $\eqref{InduceBraneUV}$, it is easy to see that the $\sqrt{-g^{\text{UV}}}$ has following expansion
\begin{align}
\label{sqrtMetUV}
&\sqrt{-g^{\text{UV}}}=1+\kappa\big(\frac{1}{2\sqrt{r_{c}}}h-\sqrt{\frac{2}{3r_{c}}}\boldsymbol{r}\big)+O(\kappa^{2})
\end{align}
Finally, by utilizing $\eqref{VeilbinSpinConnec}$ and $\eqref{sqrtMetUV}$, the effective theory localized on UV-brane is given by
\begin{align}
\nonumber
&S_{\text{UV}}=\int d^{4}x\big\{\text{i}\bar{\psi}\gamma^{\mu}\partial_{\mu}\psi-m_{e}\bar{\psi}\psi\\
\nonumber
&\quad\quad+\!\frac{\text{i}\kappa}{2\sqrt{r_{c}}}\big(\!\frac{\boldsymbol{r}}{\sqrt{6}}\bar{\psi}\gamma^{\mu}\partial_{\mu}\psi-h_{\mu}^{~\nu}\bar{\psi}\gamma^{\mu}\partial_{\nu}\psi+\partial_{\nu}h_{\mu\lambda}\bar{\psi}\eta^{\lambda[\mu}\gamma^{\nu]}\psi\big)\\
&\quad\quad+\frac{\kappa}{\sqrt{r_{c}}}(\frac{h}{2}\!-\!\sqrt{\frac{2}{3}}\boldsymbol{r})\big(\text{i}\bar{\psi}\gamma^{\mu}\partial_{\mu}\psi\!-\!m_{e}\bar{\psi}\psi\big)\big\}
\end{align}
On the other hand, the 4D effective theory should also include the contributions from the bulk geometry. In technique, the projection operation of the bulk's Riemann curvature on brane can be implemented by using the Arnowitt-Deser-Misner (ADM) formalism. In regard to perturbative metric $\eqref{PerMetric}$, the effective gravitational action on UV-brane is written as
\begin{align}
\nonumber
&S_{\text{eff-GR}}\!=\!\int d^{4}x~\big\{\frac{1}{2}h\Box h\!-\!\frac{1}{2}h^{\mu\nu}\Box h_{\mu\nu}\!-\!h\partial_{\mu}\partial_{\nu}h^{\mu\nu}\!-\!\partial_{\mu}h^{\mu\nu}\partial_{\rho}h_{~\nu}^{\rho}\\
\nonumber
&\quad\quad\quad-\frac{1}{2}m^{2}(h^{\mu\nu}h_{\mu\nu}-h^{2})-\frac{1}{2}\boldsymbol{r}\Box\boldsymbol{r}-\frac{1}{2}m^{2}\boldsymbol{r}^{2}\\
\label{EFF4DGR}
&\quad\quad\quad+\kappa\big(c_{1}\boldsymbol{r}(h^{\mu\nu}h_{\mu\nu}-h^{2})+\dots\big)\big\}
\end{align}
where the dots denote the effective operators like $O(h^3),O(h\boldsymbol{r}^2),O(h\boldsymbol{r}^3)$ which are irrelevant with the issue we focus on.

\bibliography{refs}

\end{document}